\documentclass[final,natbib,runningheads]{svjour3}

\usepackage{color,graphicx}
\usepackage{mathptmx}	   
\usepackage[T1]{fontenc}
\usepackage{latexsym}
\usepackage{textcomp}
\usepackage{ifthen}
\usepackage{verbatim}
\newboolean{manuscript}
\setboolean{manuscript}{false}
\ifthenelse{\boolean{manuscript}}{%
\usepackage[running,mathlines]{lineno}
}{%
\usepackage[running,mathlines]{lineno}
}
\usepackage{esint}
\usepackage{soul}
\usepackage{wasysym}

\usepackage{ulem}

\usepackage{subfigure}

\ifthenelse{\boolean{manuscript}}{%
}{%
\usepackage[centering,vcentering,dvips]{geometry}  
\geometry{%
papersize={15.5cm,23.5cm},
text={348pt,564pt},
headheight=40pt,
headsep=0.4\headheight,
footskip=40\lineskip}}

\usepackage{afterpage}

\usepackage[
bookmarks=true,
bookmarksnumbered=true,
hypertexnames=false,
linktocpage=true,
breaklinks=true,
colorlinks=true,
linkcolor=blue,
citecolor=blue,
filecolor=blue,
menucolor=blue,
pagecolor=blue,
urlcolor=blue,
plainpages=false,
pdfpagelabels
]{hyperref}

\smartqed  
\usepackage{etoolbox}
\newboolean{crop}
\setboolean{crop}{true}
\makeatletter
\ifthenelse{\boolean{crop}}{%
\newcommand{\includeCroppedPdf}[2][]{\begingroup%
    \edef\temp@mdfivesum{\pdf@filemdfivesum{#2.pdf}}%
    \ifcsstrequal{#2mdfivesum}{temp@mdfivesum}{}{%
      \immediate\write18{pdfcrop #2 #2-crop.pdf}}%
      \immediate\write\@auxout{\string\expandafter\string\gdef\string\csname\space #2mdfivesum\string\endcsname{\temp@mdfivesum}}%
    \includegraphics[#1]{#2-crop.pdf}\endgroup}
}{%
\newcommand{\includeCroppedPdf}[2][]{\begingroup%
    \includegraphics[#1]{#2-crop.pdf}\endgroup}
}
\makeatother

\renewcommand\cite{\citep}

\newcommand{\be}{\begin{equation}}
\newcommand{\ee}{\end{equation}}

\newcommand*{\doi}[1]{DOI \href{https://doi.org/#1}{#1}}

\makeatletter
\makeatletter

\begin{document}



\currentpdfbookmark{%
Numerical Modelling of Neutral Boundary-Layer Flow across a Forested Ridge}%
{title}
\title{%
Numerical Modelling of Neutral Boundary-Layer Flow across a Forested Ridge}

\titlerunning{%
Numerical Modelling of Neutral Boundary-Layer Flow across a Forested Ridge}

\author{%
John Tolladay~\and~%
Charles~Chemel}

\authorrunning{%
J.~Tolladay~ and C.~Chemel}

\institute{%
J.~Tolladay \at
University of Hertfordshire, College Lane, Hatfield, \mbox{AL10~9AB}, UK \\
\email{j.tolladay@herts.ac.uk}
\and
C. Chemel  \at
National Centre for Atmospheric Science (NCAS), University of Hertfordshire, UK
}

\date{%
A pre-print
}

\maketitle


\begin{abstract}
\addcontentsline{toc}{section}{Abstract}
Forest canopies have been shown to alter the dynamics of flows over complex terrain. Deficiencies have been found when tall canopies are represented in numerical simulations by an increase in roughness length at the surface. Methods of explicitly modelling a forest canopy are not commonly available in community numerical weather prediction models. In this work, such a method is applied to the community Weather Research and Forecasting model. Simulations are carried out to replicate a wind-tunnel experiment of neutral boundary-layer flow across a forested ridge. It is shown that features of the flow, such as the separated region on the lee slope of the ridge, are reproduced by the roughness length or canopy model methods. Shear at the top of the ridge generates turbulence that spreads vertically as the flow moves downstream in both cases, but is elevated to canopy top where a canopy model is used. The roughness-length approach is shown to suffer several deficiencies, such as an over-prediction of mean wind-speeds, a lack of turbulence over flat forested ground and an insufficient vertical extent of turbulence at all locations of the domain studied. Sensitivity to the horizontal resolution of the simulation is explored. It is found that higher resolution simulations improve reproduction of the mean flow when modelling the canopy explicitly. However, higher resolutions do not provide improvements for the roughness-length case and lead to a reduction in the horizontal extent of the separated region of flow on the lee slope of the ridge.
\end{abstract}


\section{Introduction}

Interactions with surface elements such as buildings and forested areas have a significant effect on the flows present in complex (uneven) terrain \citep{fernando10}. It is thereby important to understand the contributing factors to these flows, so as to evaluate their impacts on pooling of cold air and air pollution in valleys or to predict mean wind speed and turbulence statistics for wind farm applications. \citet{bastin19} estimated that $2.8$ of the $15$ billion hectares (that is $18.7$\%) of the Earth's land surface are covered in a forest canopy with a tree cover greater than $10$\%. Given the difficulty of building on ridge or valley sides and in mountainous areas, these areas are often left untouched and so likely to be covered in shrubs and trees at mid-latitudes. Therefore, a significant fraction of complex terrain is likely to be covered in a forest canopy of some sort.

\par
\citet{finnigan00} reviewed the bulk of the work done to date to understand flow in homogeneous forest canopies over flat ground. \citet{belcher12} built on this review to illustrate how canopy flows respond to complex terrain. More recently, \citet{finnigan20} reviewed the subject of boundary layer flows in complex terrain. A section of this review focused on theory, analytical and numerical models of flow over canopy covered hills, and considered the effects of stability and scalar transport. There is a reasonably good mechanistic understanding of the dynamics of canopy flows, the adjustment of flows at canopy edges, the ability to use `simple' turbulence closures due to the inviscid nature of the dynamics, effects of forested terrain on scalar transport, the generation of reversed flows within canopies downstream of ridge-tops and the significant reduction of turbulence and momentum within a canopy. However, woodland canopies are often not modelled explicitly in numerical weather prediction (NWP) models and relatively little attention has been paid to the evaluation of the effects of forest cover in complex terrain in these models.

\par
Due to the broad range of scales of canopy elements (e.g.~leaves, twigs and branches), it is currently computationally impractical to model explicitly the processes involved in the canopy flow dynamics on the scale of canopy elements in NWP models. Recognising the increased friction caused by canopy elements, the most common approach to parametrise the effects of the canopy on the flow is to increase the roughness $z_{0}$ of the underlying surface and displace the height of the ground. As is customary in micrometeorology, let us align the $x$-direction with that of the mean horizontal flow (i.e.~the stream-wise direction) of velocity~$\langle{u}\rangle$ such that there is no variation of the mean span-wise component of velocity~$\langle{v}\rangle$ in the $y$-direction. Parametrising the turbulent kinematic flux of stream-wise momentum in the vertical direction $z$ (referred to as stream-wise momentum flux thereafter), $\langle{u'w'}\rangle$, using a first-order flux--gradient relationship yields
\be
u_{\star}^{2}\equiv -\langle{u'w'}\rangle=K_{m}\,\frac{\partial{\langle{u}\rangle}}{\partial{z}}
\mbox{,}
\label{eq:K_theory}
\ee
where $u_{\star}\equiv\sqrt{\left|{\langle{u'w'}\rangle}\right|}$ is the (stream-wise) friction velocity and $K_{m}$ is the eddy diffusivity of momentum. Using a mixing-length model, $K_{m}$ is modelled, for neutral stability, as
\be
K_{m}=\ell\,u_{\star}=\ell^{2}\,\frac{\partial{\langle{u}\rangle}}{\partial{z}}
\mbox{,}
\ee
where the mixing-length $\ell=\kappa\,\left({z-d}\right)$, $\kappa=0.4$ is the von~K\'{a}rm\'{a}n constant and $d$ is the zero-plane displacement height, where $\langle{u}\rangle=0$. Assuming that $\langle{u'w'}\rangle$ is constant in the surface layer (that is $u_{\star}$ is constant in this `constant-flux' layer), an integration of Equation~(\ref{eq:K_theory}) with the boundary condition $\langle{u}\rangle\left({z=d+z_{0}}\right)=0$ gives the logarithmic law
\be
\langle{u}\rangle=\frac{u_{\star{0}}}{\kappa}\,\ln\left(\frac{z-d}{z_{0}}\right)
\mbox{,}
\label{eq:loglaw}
\ee
where $u_{\star{0}}=u_{\star}\left({z=0}\right)$. Since the wind speed is reduced to zero at the displacement height, this formulation cannot represent the flow or turbulence below the displacement height. When canopy elements are small in scale relative to the extent of the atmosphere being modelled above, flows within the canopy have little impact on the dynamics of the flow well above the canopy. A change in roughness length at the surface is, thereby, a reasonable option to be used in NWP models for flows over short canopies. However, when dealing with taller canopy elements on the scales of mature trees, in the range $10$--$50$\,m in height, turbulence and drag within the canopy can have a profound effect on the flow above canopy \citep[e.g.][]{ross12}, which is not accounted for over a bare surface with increased roughness. Furthermore, the mixing length $\ell$ as defined above is not the most appropriate to use as a length scale for turbulent motions within a forest canopy \citep{wilson98}.

\par
Several studies using the roughness-length approach to simulate the flow over forested ridges have shown that various features of the flow are not recreated accurately \cite[e.g.][]{finnigan95,ross05}. In particular, the region of separated flow, where the wind downstream of a forested ridge reverses close to the ground, is often not as substantial in simulations using a roughness-length parametrisation when compared to observations. The roughness-length approach also tends to over-predict the turbulence kinetic energy (TKE) within close proximity to ridge-tops. \citet{finnigan04} and \citet{harman07} examined canopy flows with an analytical model and found evidence that canopies do not tend to a constant roughness length while flowing over hills. This suggests that using a constant roughness length across a forested section of a hill is unlikely to properly recreate the flows over forested, complex terrain. A study by \citet{allen06}, on the effects of roughness lengths on flows over ridges, found that flow separation is encouraged if the roughness length is largest at the top of ridges. If the surface roughness length is largest at the base of ridges then flow separation is reduced.

\par
More success has been found in recreating flow dynamics within and above canopies where the effects of these canopies are modelled explicitly and with a proper vertical extent. Such models consider the canopy as a horizontally homogeneous but vertically resolved volume, wherein the total kinematic drag generated by the canopy, $\mathbf{F}_{c}$, is expressed as the product of a drag coefficient $C_{d}$, a one-sided plant area density $a$ and the square of the resolved velocity $\mathbf{u}$ \citep[see][]{wilson77,raupach81,raupach82,finnigan85,raupach86}, namely
\be
\mathbf{F}_{c}=-C_{d}\,a\,\left|{\mathbf{u}}\right|\,\mathbf{u}
\mbox{.}
\label{eq:canopy_drag}
\ee
This term represents the momentum per unit mass that is lost per unit time through interaction between the air and the trunks, branches and leaves that make up a forest canopy. Numerous studies added $\mathbf{F}_{c}$ as an additional sink of momentum to the momentum equation within the canopy in bespoke numerical models to study {\it (i)} observed properties of the flow within the canopy over flat ground, e.g.~sweeps and ejections that govern turbulent transport \citep[e.g.][]{shaw92,dupont08b,finnigan09,ouwersloot17}, effects of forest edges \citep[e.g.][]{cassiani08,dupont11}, variability in plant-area density in the horizontal \citep[e.g.][]{bohrer09} and in the vertical \citep[e.g.][]{dupont08a}, and {\it (ii)} the associated biosphere--atmosphere exchange of trace-gas and other scalars \citep[e.g.][]{patton01,patton03}. In the study by \citet{ouwersloot17}, some work was done to assess the sensitivity of model results to changes in grid resolution, but this related only to the influence of resolution on the production of unphysical velocity fluctuations caused by the sharp transition in plant area density at canopy top.

\par
Less attention was given to deep canopies over hilly and mountainous terrain. Numerical modelling studies of flow across forested hills and ridges were conducted to challenge the model results with experimental measurements or analytical predictions \citep{ross05,tamura07,dupont08b,ross08,grant16}. \citet{patton09} investigated phase relationships between mean flow variables and turbulence statistics and their sensitivity to a change in leaf area density. The separation region was found to be ambiguous for sparse canopies while being well-defined within the canopy on the lee side of the ridge for dense canopies. \citet{ross11} and \citet{chen19} examined how topography-induced changes in the flow translate to scalar transport within the canopy. Scalars emitted near the ground exhibited larger spatial variability than those emitted in the upper canopy \citep[see also][]{ross15}. Transport out of the canopy was enhanced compared to that over flat terrain, with a preferential route out of the canopy located over the region of separated flow. \citet{ross13} considered the effects of a forest canopy covering partially hilly terrain. Flow separation was essentially limited to the forested region over the lee slope where an adverse pressure gradient is induced by the terrain. The differences in flow separation for different positionings of the forest were found to have a large impact of scalar transport out of the canopy.

\par
In the present work, a simple canopy model is implemented in the Weather Research and Forecasting model. Large-eddy simulations using the standard roughness length approach and using the canopy model are evaluated using the `Furry Hill' data \citep{finnigan95} for a neutral boundary-layer flow across a forested ridge. Simulations are carried out using a range of horizontal and vertical grid spacings in order to assess the impact that this has on the response of the flow to the canopy covered hill, hereafter referred to as a ridge to clarify the two dimensional profile. The modelling system is presented briefly in Sect.~\ref{sec:canopy_model}. The set-up of the modelling system and the design of the numerical experiments are described in Sect.~\ref{sec:model_setup}. Numerical results and sensitivities to the horizontal grid spacing are analysed in Sect.~\ref{sec:results}. Conclusions are given in Sect.~\ref{sec:conc}.


\section{Modelling System}
\label{sec:canopy_model}

Numerical simulations were performed with the community Weather Research and Forecasting (WRF) modelling system, version $3.9.1$, and using its Advanced Research WRF (ARW) dynamical core. The ARW dynamical core integrates the fully compressible, non-hydrostatic equations of motion in flux form. The equations are discretised using a terrain-following mass-based coordinate system and a staggered grid of type Arakawa-C. Time integration was performed using a third-order Runge-Kutta scheme and a time-splitting technique with semi-implicit sound waves. A fifth-order Weighted Essentially Non-Oscillatory (WENO) scheme with a positive definite filter was selected for advection of momentum and scalar variables. The Coriolis force was excluded as the effects of the Earth's rotation on the flow are negligible at the scales of motion that are considered in the present work (see Sect.~\ref{sec:model_setup}).

\par
Canopy models have been developed and implemented in WRF by, for example, \citet{ma19} and \citet{arthur19}. However, these were not finalised until considerable work had already been done for this study. The term $\mathbf{F}_{c}$, defined by Equation~(\ref{eq:canopy_drag}), as an additional sink of momentum to the momentum equation within the canopy was therefore implemented by the authors. The $1.5$-order turbulence closure scheme developed by \citet{deardorff80} with a prognostic equation for sub-grid-scale (SGS) TKE, denoted by $k_{\mbox{\scriptsize SGS}}$, was used to determine the SGS fluxes from the resolved fields and the SGS TKE. In this scheme, the eddy viscosity of momentum $K_{m}$ is modelled as
\be
K_{m}=C_{k}\,\lambda\,\sqrt{k_{\mbox{\scriptsize SGS}}}
\mbox{,}
\ee
where the diffusion coefficient $C_{k}=0.10$ and the SGS mixing length scale $\lambda$ was set equal to the cube root of the grid-cell volume $\Delta{s}=({{\rm\Delta{x}}\,{\rm\Delta{y}}\, {\rm\Delta{z}}})^{1/3}$.

\par
In the turbulence closure scheme proposed by \citet{deardorff80} the SGS TKE dissipation is given by
\be
\epsilon_{v}=C_{\epsilon}\,k_{\mbox{\scriptsize SGS}}^{3/2}/\lambda
\mbox{,}
\ee
where the dissipation coefficient $C_{\epsilon}=0.93$, except in the first grid cell immediately above the surface, where $C_{\epsilon}$ is increased to $3.9$ to mimic a `wall effect' so as to prevent $k_{\mbox{\scriptsize SGS}}$ from becoming unduly large there. Following \citet{shaw92}, the standard viscous dissipation $\epsilon_{v}$ was augmented by an additional dissipation term,
\be
\epsilon_{c}=2\,C_{d}\,a\left|{\mathbf{u}}\right|\,k_{\mbox{\scriptsize SGS}}
\mbox{,}
\ee
to represent the dissipation caused by the interactions between the air and the canopy.


\section{Design of The Numerical Experiments}
\label{sec:model_setup}
\begin{figure*}
{\scalebox{1.0}{\includeCroppedPdf[width=1.0\linewidth]{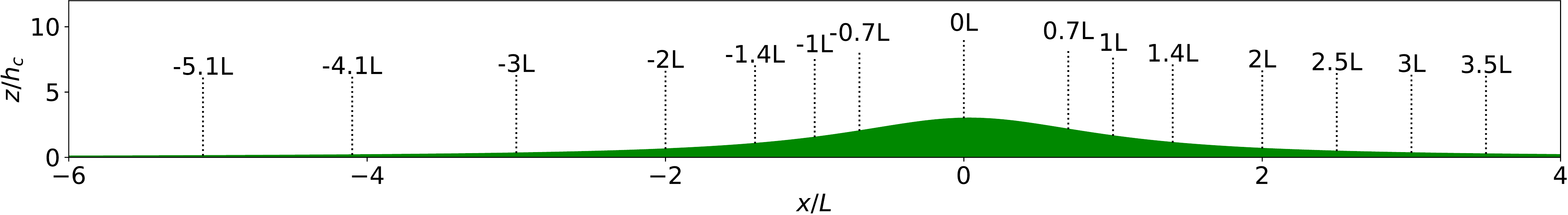}}}
\caption{Terrain height $z_{g}$, normalised by the height of the canopy $h_{c}$, along the stream-wise direction $x$. The distance along $x$ is normalised by the half-height width $L$ of the ridge. The terrain is symmetric about $x=0$ and uniform in the span-wise direction $y$ (into the page). The vertical dotted lines indicate the positions where experimental data is available}
\label{fig:guide}
\end{figure*}

The `Furry Hill' wind-tunnel experiment carried out by \citet{finnigan95} is used to evaluate the different methods of parametrising the canopy presented in the previous sections. In this experiment, a neutral atmosphere with a uniform background wind $U_{b}=12$\,m\,s$^{-1}$ interacts with a forested ridge with a two dimensional profile of a `witch of Agnesi' centred about $x=0$ (see~Fig.~\ref{fig:guide}). Ground level $z_{g}$ is defined as $z_{g}=H_{e}/[1+(x/L_{e})^{2}]$, where the height of the ridge $H_{e}=0.15$\,m and its half-height width $L_{e}=0.42$\,m. For the artificial canopy used in the experiment, the height of the canopy was $h_{c,e}=0.047$\,m, the plant area density $a_{e}=10$\,m$^{-1}$ and the drag coefficient $C_{d,e}=0.68$, leading to a canopy-drag length scale $L_{c,e}=(C_{d,e}\,a_{e})^{-1}=0.147$\,m. This artificial canopy was then surrounded by a rough surface of gravel with a diameter of $0.014$\,m ($\approx h_{c,e}/3.36$).


\par
The physical properties of the ridge and canopy are scaled up as proposed by \citet{dupont08b}, so that the experiment represents atmospheric scales. This provides a ridge height $H=30$\,m, half-height width $L=84$\,m, canopy height $h_{c}=10$\,m, plant area density $a=0.165$\,m$^{-1}$ and drag parameter $C_{d}=0.2$, leading to a canopy-drag length scale $L_{c}=(C_{d}\,a)^{-1}=30.3$\,m. Geometric similarity is achieved ($H/L=H_{e}/L_{e}=0.36$), but $a$ was increased slightly in comparison to the value of $0.16$\,m$^{-1}$ used by \citet{dupont08b}. This was done to achieve marginally closer values to the `Furry Hill' experiment for the conditions $L_{c,e}/L_{e}=0.35$ and $h_{c,e}/L_{c,e}=0.32$, where these are $L_{c}/L=0.36$ and $h_{c}/L_{c}=0.33$ for the values used here. The canopy extends from $x/L=-9.35$ to $3.39$ to match the location of the artificial canopy used in the experiment. For all simulations, the surrounding gravel surface is simulated using a roughness length of $z_{0}=0.0025\,h_c=0.025$\,m, approximately one tenth of the diameter of the scaled up gravel.

\par
Simulations are performed with two nested domains and feedback was enabled, such that the lateral boundaries of the inner domain are set by the outer domain solution and the solution at the boundaries of the inner domain are fed back to the outer domain. The outer domain extends horizontally between $x/L=\pm 36$ and $y/L=\pm 10.6$. The inner domain, centred on $(x,y)=(0,0)$, covers one third of this extent, between $x/L=\pm 12$ and $y/L=\pm 3.53$. It should be noted that the ridge and canopy are present in the inner and outer domains and both domains are run in large-eddy simulation (LES) mode. In the vertical, a terrain-following coordinate is used with $85$ points from the surface to the top of the two domains at approximately $z/h_c=20$ above the sections of flat ground. A hyperbolic tangent function is applied to the vertical grid spacings to compact the grid close to the ground, with grid spacings barely increasing above. The grid is stretched such that the lowest level has a height of approximately $h/h_c=0.1$, providing $10$ levels within the canopy. A lowest grid level height of $h/h_c=0.2$ was also considered but the results are not shown because the difference between the two cases was negligible. The top of the two domains is frictionless and includes a $50$\,m deep Rayleigh damping layer to reduce numerical instabilities in the simulation \citep{klemp08}. Periodic boundary conditions are used at the lateral boundaries of the outer domain, which is given a sufficient extent in the stream-wise direction for the forested ridge to have no noticeable effect on the incoming flow at the inner domain. Horizontal grid spacings ${\rm\Delta}{x}={\rm\Delta}{y}$ of $0.024$, $0.048$ and $0.071\,L$ ($2$, $4$ and $6$\,m) are considered for the inner domain and $0.071$, $0.143$ and $0.214\,L$ ($6$, $12$ and $18$\,m) for the outer domain, respectively. At the surface, no-slip conditions are imposed, the heat flux is set to zero and the momentum flux is calculated from wind velocity at the first grid point above the surface, using the logarithmic profile of Eq.~(\ref{eq:loglaw}) with a prescribed surface roughness length $z_{0}$.

\par
\begin{figure*}
\centering{\scalebox{1.0}{\includeCroppedPdf[width=1.0\linewidth]{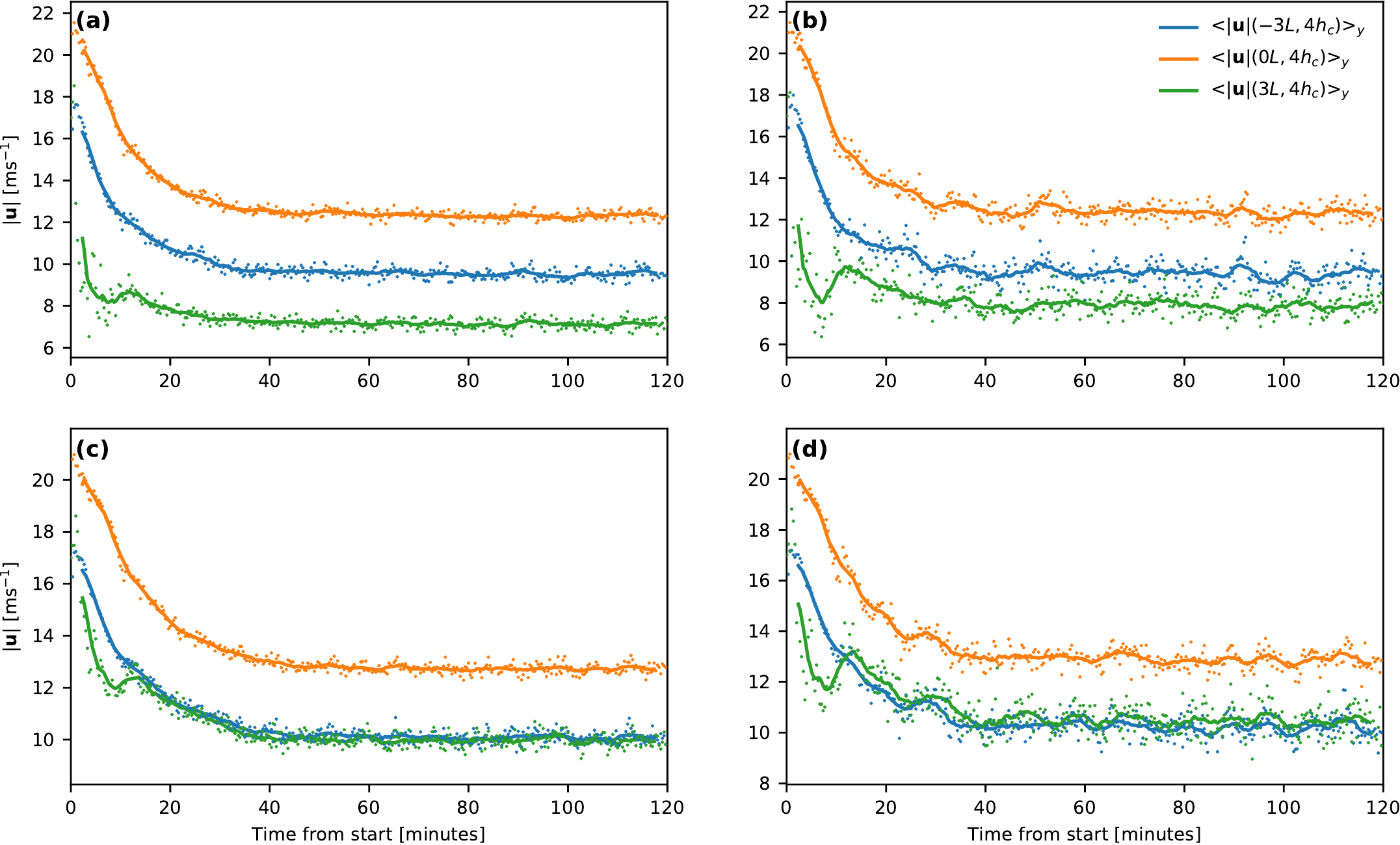}}}
\caption{Wind speed $\left|{\mathbf{u}}\right|$ averaged in the $y$-direction for \mbox{WRF-C6} {\bf (a)} outer and {\bf (b)} inner domains and for \mbox{WRF-R6} {\bf (c)} outer and {\bf (d)} inner domains for positions around the ridge at $4\,h_c$ above ground level. The points show instantaneous wind speeds, while the solid lines show a $5$\,min moving average}
\label{fig:stability}
\end{figure*}

The simulations are initialised with a wind speed of $17$\,m\,s$^{-1}$ in the positive $x$-direction at all positions. The initial wind speed is larger than that of the upstream flow in the wind-tunnel experiment ($12$\,m\,s$^{-1}$) so that the flow speed achieved after the spin-up period corresponds to that approaching the canopy in the experiment. After approximately $30$ to $40$\,minutes of simulation time, the solution reaches a quasi-steady state, as shown in Fig.~\ref{fig:stability} by the relatively constant wind speed after this time. The $5$\,minute moving average of wind speed at the locations shown vary by no more than $3$\% in the outer domain and $6$\% in the inner domain for the time period after the first $45$\,minutes for all the simulations that were performed. To obtain numerically stable results, the vertical grid resolution and maximum flow speed demanded a model time-step ${\rm\Delta}{t}=0.025$\,s, with data being exported at $20$\,s intervals. The $90$ data points between minute $45$ and minute $75$ of the simulation are then used to calculate the mean velocity components $\langle{u}\rangle_{yt}$, $\langle{v}\rangle_{yt}$, $\langle{w}\rangle_{yt}$ and their variances $\langle{u'^2}\rangle_{yt}$, $\langle{v'^2}\rangle_{yt}$, $\langle{w'^2}\rangle_{yt}$ in the $x$-, $y$-, $z$-directions, respectively, the momentum flux per unit mass $\langle{u'w'}\rangle_{yt}$ and TKE per unit mass $\langle{k}\rangle_{yt}$, where $\langle{\Box}\rangle_{yt}$ denotes an average in both time and the span-wise direction $y$ and $'$ represents a fluctuation from this averaged value. The velocity components and turbulence statistics collected by \citet{finnigan95} were normalised with the friction velocity $u_\star$ at canopy top at $x/L=-3$. In this work the results are presented in SI units and the normalisation of the measurements was reversed using $u_\star=0.911$\,ms$^{-1}$ calculated from their results.

\par
In the following, the WRF simulations that use the canopy model introduced in Sect.~\ref{sec:canopy_model} are given the reference \mbox{WRF-C}. The surface roughness length at the location of the canopy is set to $z_{0}=0.001\,h_{c}=0.01$\,m \citep[as in][]{shaw92} for these simulations. Another set of WRF simulations with reference \mbox{WRF-R} use only a change in surface roughness length to represent the artificial canopy, rather than modelling it explicitly. For \mbox{WRF-R} the roughness length used at the location of the canopy is $z_{0}=0.085\,h_c=0.85$\,m, determined to provide the closest agreement with the profile above the flat section of canopy in \mbox{WRF-C} when used in Equation~(\ref{eq:loglaw}). When discussing a specific simulation, the references \mbox{WRF-C} and \mbox{WRF-R} are followed by a number representing the horizontal grid spacing of the inner domain. For example, with horizontal grid spacing of $2$\,m ($0.024\,L$) for a simulation using the explicit canopy model, the reference would be \mbox{WRF-C2}. The height of the lowest grid level is below the displacement height of the canopy and WRF is designed to only apply the effects of a roughness length in the bottom grid level. The results for \mbox{WRF-R} were elevated upwards by $z/h_c=0.6$ such that the displacement height $d=0.7\,h_{c}$ is within the lowest grid level. To keep notation simple in the following, height above ground level $h=z-z_{g}$ for \mbox{WRF-R} refers to that for \mbox{WRF-C}. Note that the ground is elevated at all positions, as this change in displacement height can not be properly applied at the leading and trailing edge of the canopy-covered region.


\begin{figure*}
\centering{\scalebox{1.0}{\includeCroppedPdf[width=1.0\linewidth]{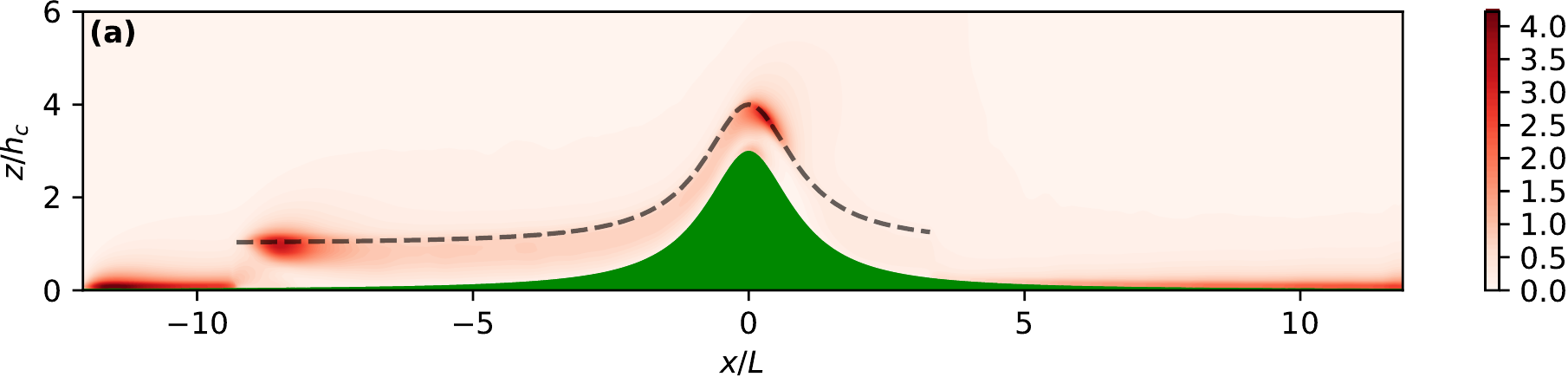}}}
\centering{\scalebox{1.0}{\includeCroppedPdf[width=1.0\linewidth]{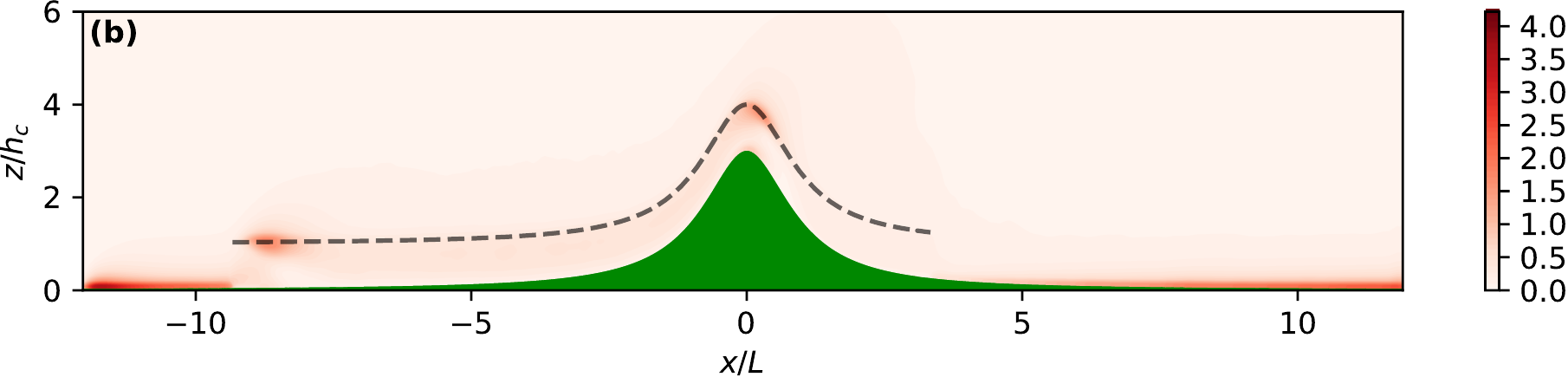}}}
\centering{\scalebox{1.0}{\includeCroppedPdf[width=1.0\linewidth]{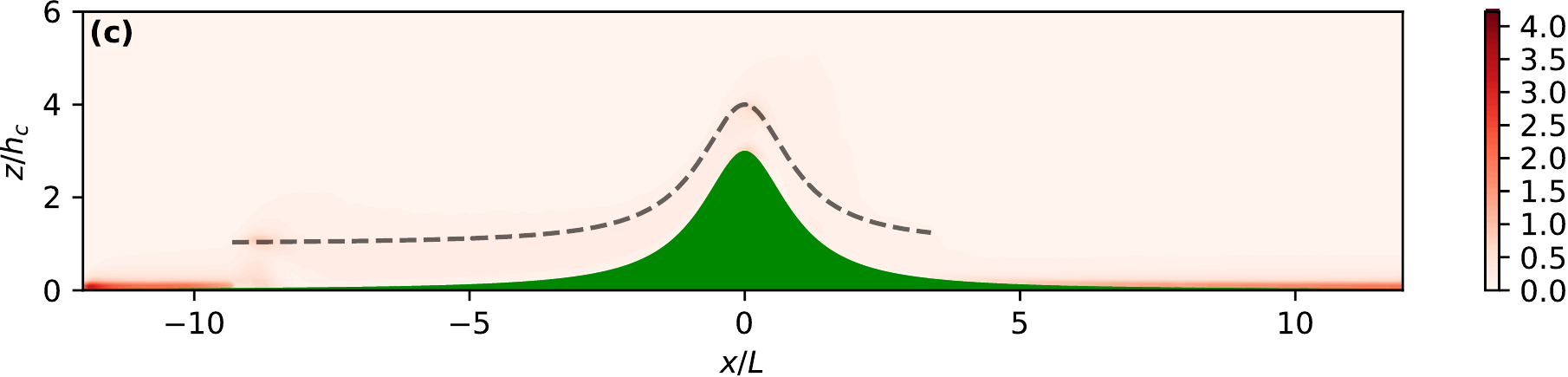}}}
\caption{Ratio of sub-grid scale to resolved turbulence kinetic energy per unit mass, $\langle{k_{\mbox{\scriptsize SGS}}}\rangle_{yt}/\langle{k}\rangle_{yt}$ for {\bf (a)} \mbox{WRF-C6}, {\bf (b)} \mbox{WRF-C4} and {\bf (c)} \mbox{WRF-C2}. The top of the simulated artificial canopy is indicated by a black dashed line}
\label{fig:cross_k_ratio}
\end{figure*}


\par
To ascertain that the model does resolve the most energetic scales of motion, the ratio of SGS to resolved TKE is shown in Fig.~\ref{fig:cross_k_ratio} for \mbox{WRF-C}. The SGS component of $\langle{k}\rangle_{yt}$ is negligible at all positions away from the canopy regardless of the horizontal grid spacing considered. In all cases, SGS TKE is up to $4$ times greater in magnitude than resolved TKE at the lowest levels of the domain, outside of the canopy. While it is not shown here, the same is true for all cases of \mbox{WRF-R} at the lowest grid levels but with a more significant contribution where the surface roughness increases to represent the canopy, especially around the peak of the ridge between $x/L=-1$ to $1$. For \mbox{WRF-C6} [see Fig.~\ref{fig:cross_k_ratio}(a)], above the lowest grid levels, SGS TKE is most substantial (up to $3$ times resolved TKE) at the top of the leading edge of the canopy ($x/L=-9$ to $-8$) and at canopy top near the peak of the ridge ($x/L=0$ to $1$). As the horizontal grid spacing is reduced, more of the turbulence generated in these locations within the canopy is resolved. Almost all TKE is resolved for \mbox{WRF-C2} [see Fig.~\ref{fig:cross_k_ratio}(c)]. It is therefore expected that the results for \mbox{WRF-C2} will be closer to the measurements made in the `Furry Hill' experiment than the other cases of \mbox{WRF-C}.


\section{Results and Analysis}
\label{sec:results}

%
%

\subsection{Model Evaluation}
\label{subsec:evaluation}

\begin{figure*}
\centering{\scalebox{1.0}{\includeCroppedPdf[width=1.0\linewidth]{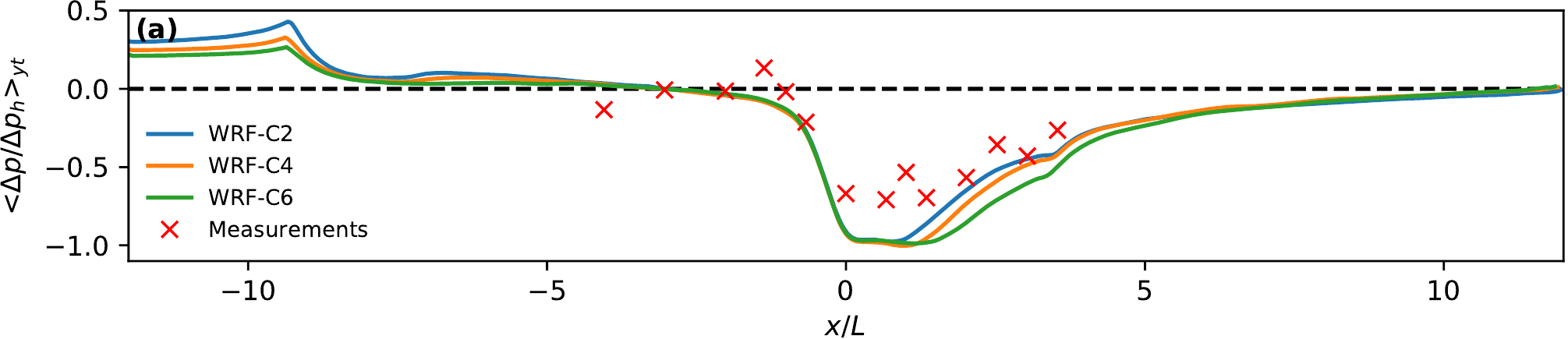}}}
\centering{\scalebox{1.0}{\includeCroppedPdf[width=1.0\linewidth]{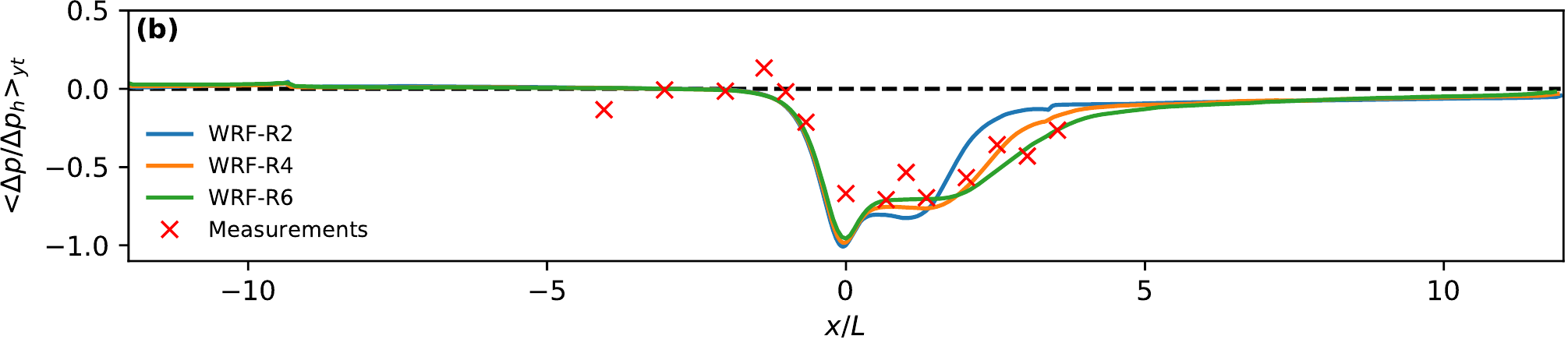}}}
\caption{Simulated mean near-surface normalised pressure $\langle{\Delta{p}/\Delta{p}_{h}}\rangle_{yt}$ for {\bf (a)} \mbox{WRF-C} and {\bf (b)} \mbox{WRF-R} compared with the wind-tunnel measurements over the `Furry Hill' from \citet{finnigan95} (see text for details). The pressure perturbation ${\rm\Delta}{p}$ is calculated relative to the value at $x/L=-3$}
\label{fig:pressure1}
\end{figure*}

Pressure perturbations $\langle{\Delta{p}}\rangle_{yt}$ as a difference from the values at corresponding altitude $z$ further upstream, taken here at $x/L=-3$, result from interaction with the forested ridge. The pressure perturbations induced by the canopy, $\Delta{p}_{c}$, and by the ridge,~$\Delta{p}_{h}$, scale as $\Delta{p}_{c}\approx\rho\,U_{b}^{2}\,h_{c}/L_{c}$ and $\Delta{p}_{h}\approx\rho\,U_{b}^{2}\,H/L$, respectively, where $\rho$ is density and $U_b$ is the stream-wise velocity at corresponding altitude upstream of the ridge, taken here at $x/L=-5$ \citep{belcher03}. Simulated near-surface pressure perturbation for \mbox{WRF-C} and \mbox{WRF-R} is compared with measurements from the wind-tunnel experiment in Fig.~\ref{fig:pressure1}(a) and (b), respectively. Recall that the simulated fields for \mbox{WRF-R} are valid from height $z=d$ and so pressure is compared at this height. The simulations capture reasonably well the measured decrease in pressure across the top of the ridge, although both over-predict the drop in pressure over the windward slope of the ridge. Immediately after the ridge-top the near-surface pressure for \mbox{WRF-R} follows the measurements closely, with the exception of \mbox{WRF-R2} where the near-surface pressure increases over a shorter distance than for the other \mbox{WRF-R} simulations. For \mbox{WRF-C}, the distribution of near-surface pressure is similar to that measured but of a larger magnitude at all positions where measurements are available. This shows, as pointed out for instance by \citet{ross05}, that the effective width of the ridge is increased for \mbox{WRF-C} when compared to \mbox{WRF-R}.

\begin{figure*}
\centering{\scalebox{1.0}{\hspace{0.0451\linewidth} \includeCroppedPdf[width=0.947\linewidth]{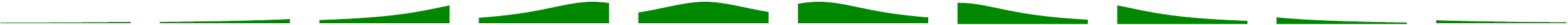} \hspace{0.007\linewidth}}}
\centering{\scalebox{1.0}{\includeCroppedPdf[width=1.0\linewidth]{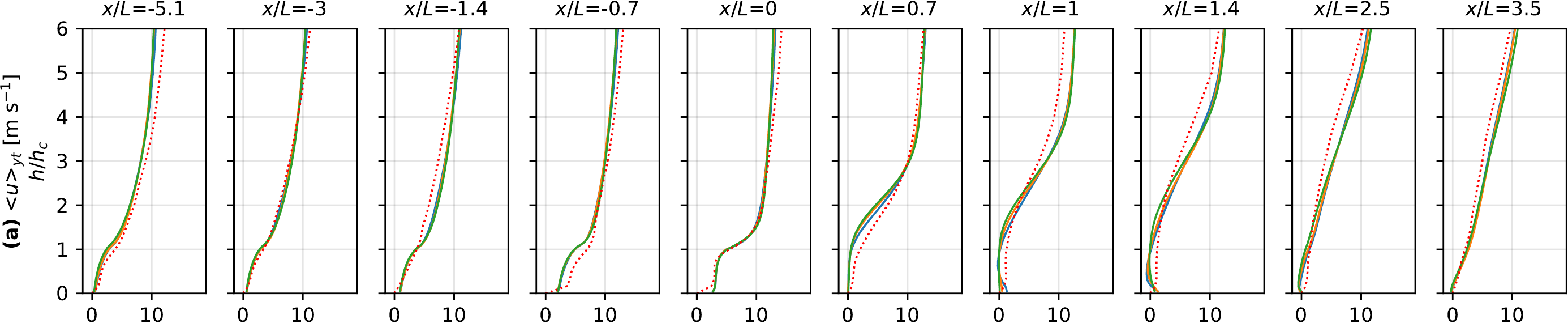}}}
\centering{\scalebox{1.0}{\includeCroppedPdf[width=1.0\linewidth]{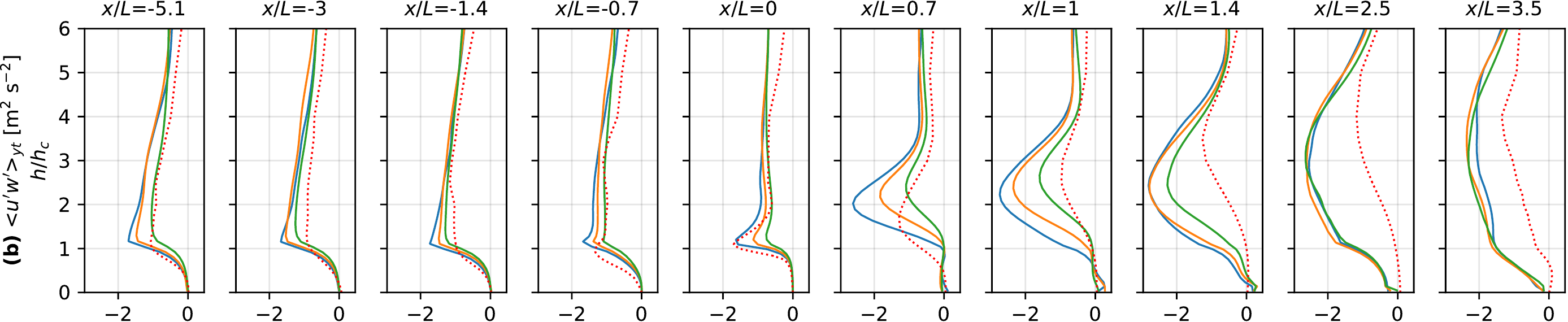}}}
\centering{\scalebox{1.0}{\includeCroppedPdf[width=1.0\linewidth]{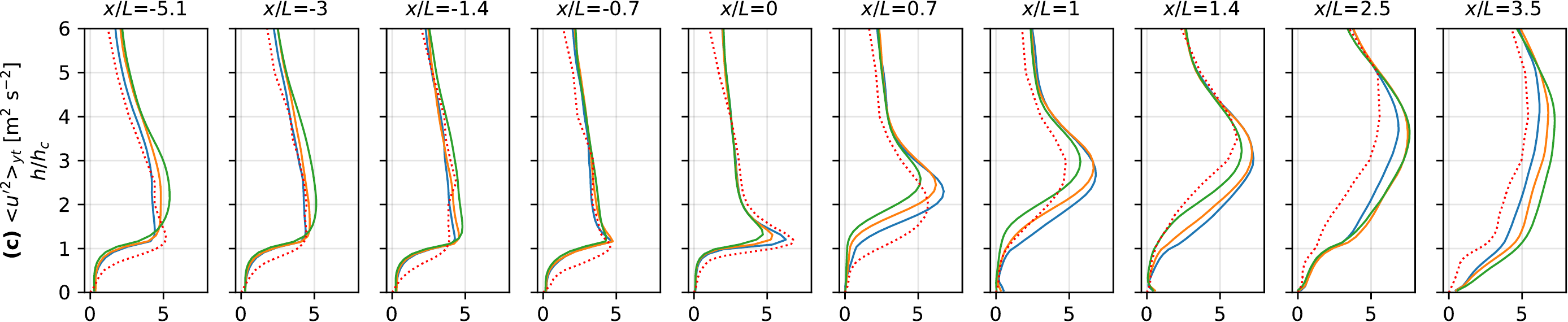}}}
\centering{\scalebox{1.0}{\includeCroppedPdf[width=1.0\linewidth]{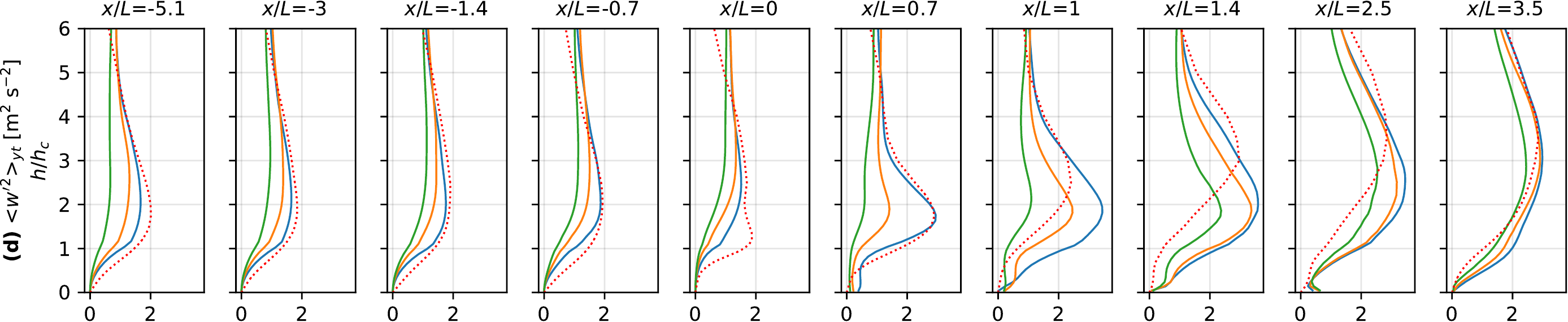}}}
\centering{\scalebox{1.0}{\includeCroppedPdf[width=1.0\linewidth]{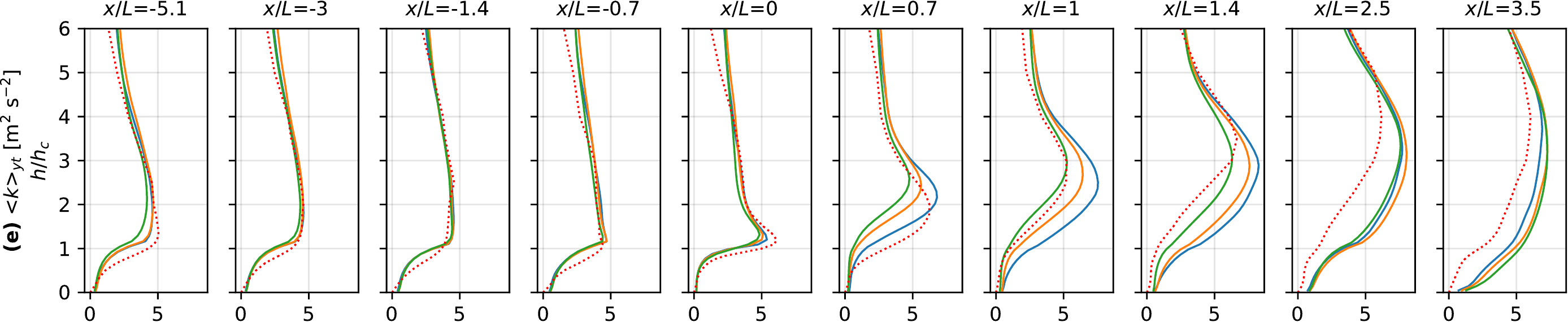}}}
\centering{\scalebox{1.0}{\includeCroppedPdf[width=0.5\linewidth]{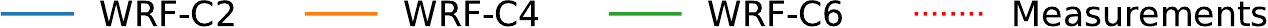}}}
\caption{Vertical profiles of time and $y$-direction averaged (from the top to bottom panels) stream-wise velocity $\langle{u}\rangle_{yt}$, vertical momentum flux per unit mass $\langle{u'w'}\rangle_{yt}$, stream-wise velocity variance $\langle{u'^{2}}\rangle_{yt}$, vertical velocity variance $\langle{w'^{2}}\rangle_{yt}$ and turbulence kinetic energy per unit mass $\langle{k}\rangle_{yt}$ for \mbox{WRF-C}, compared with the wind-tunnel measurements over the `Furry Hill' from \citet{finnigan95} (see text for details). A graphical representation of the slope in the terrain present around each profile is displayed at the top of the figure}
\label{fig:profiles_can}
\end{figure*}

\begin{figure*}
\centering{\scalebox{1.0}{\hspace{0.0451\linewidth} \includeCroppedPdf[width=0.947\linewidth]{profile-slope-markers} \hspace{0.007\linewidth}}}
\centering{\scalebox{1.0}{\includeCroppedPdf[width=1.0\linewidth]{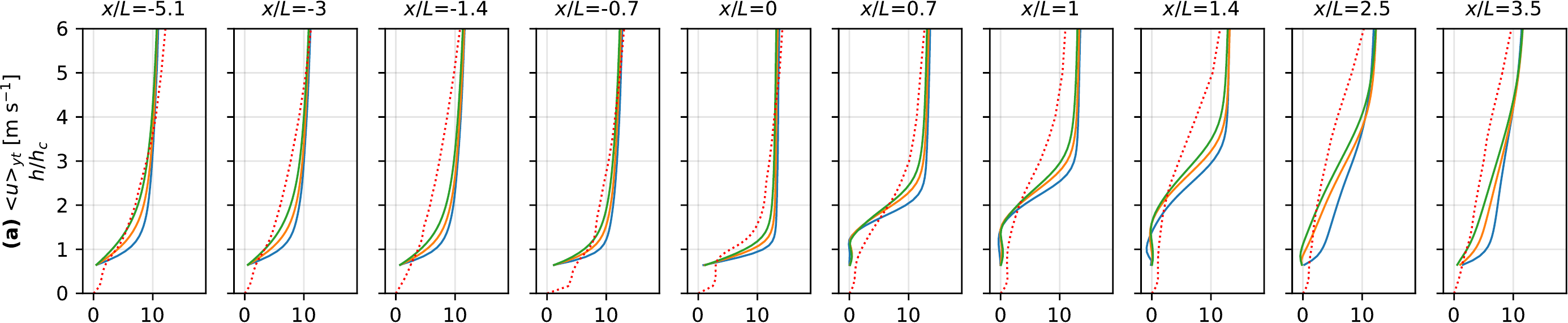}}}
\centering{\scalebox{1.0}{\includeCroppedPdf[width=1.0\linewidth]{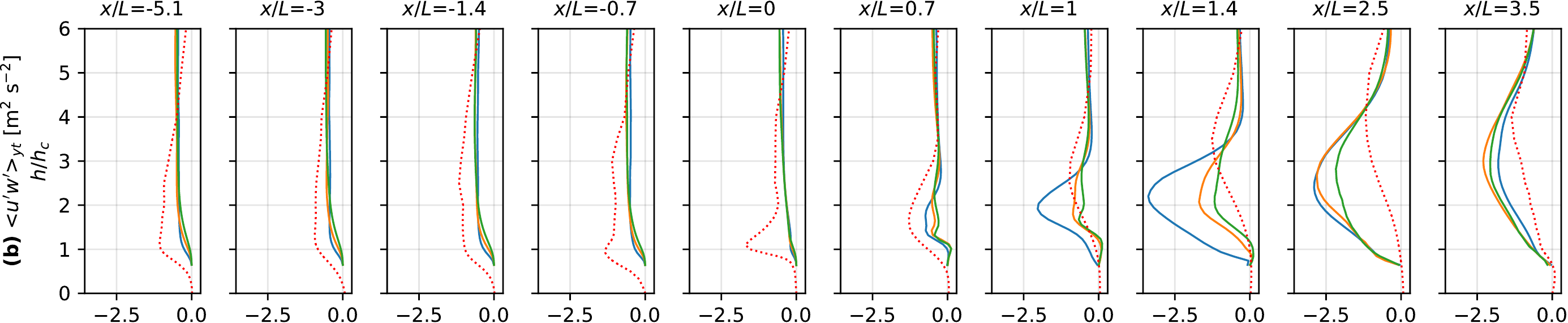}}}
\centering{\scalebox{1.0}{\includeCroppedPdf[width=1.0\linewidth]{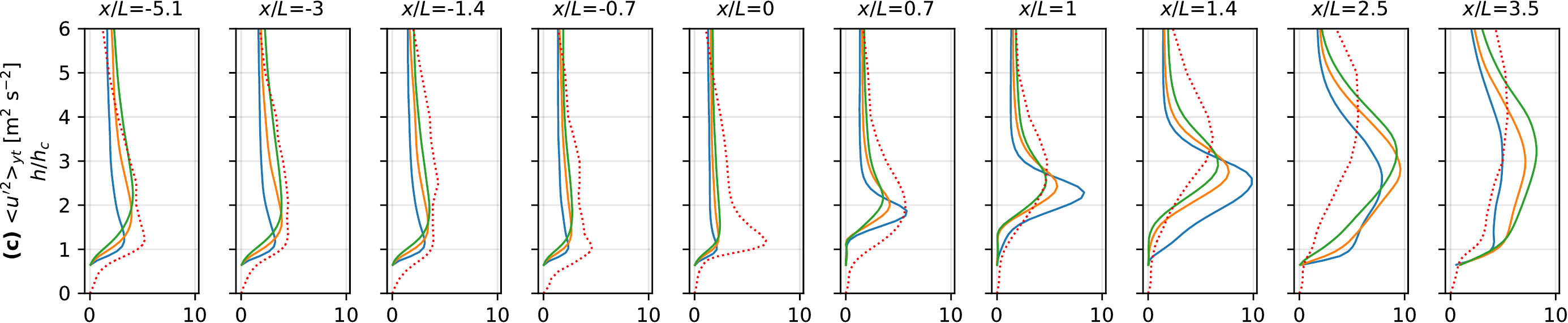}}}
\centering{\scalebox{1.0}{\includeCroppedPdf[width=1.0\linewidth]{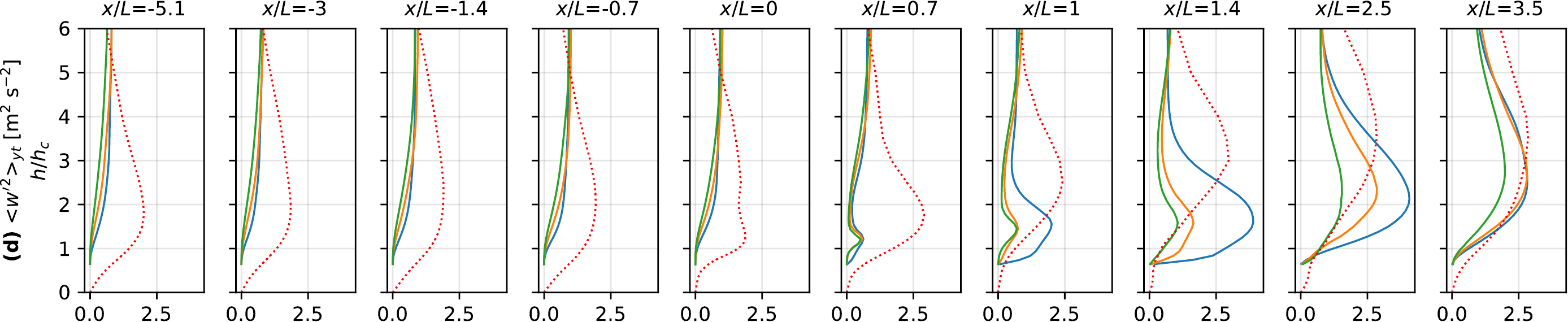}}}
\centering{\scalebox{1.0}{\includeCroppedPdf[width=1.0\linewidth]{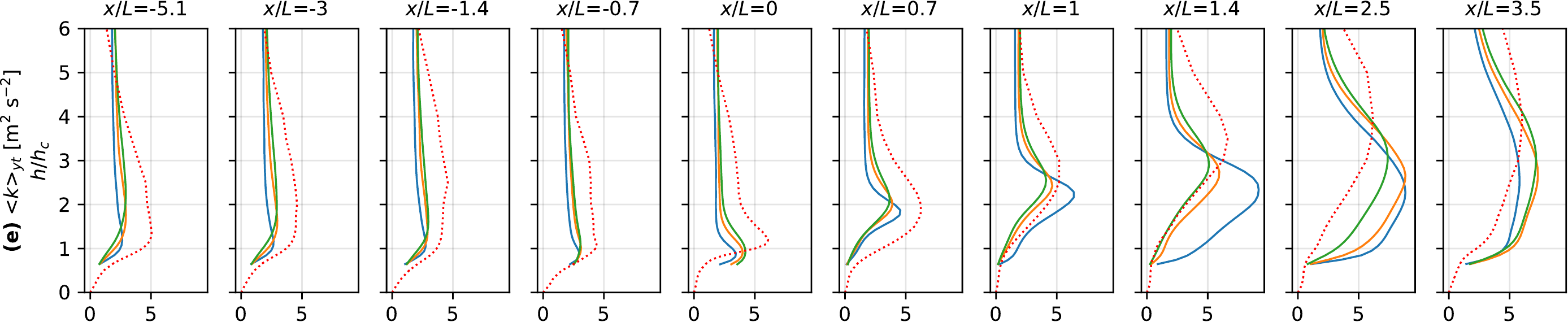}}}
\centering{\scalebox{1.0}{\includeCroppedPdf[width=0.5\linewidth]{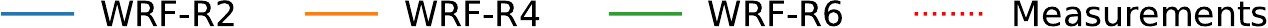}}}
\caption{Vertical profiles of time and $y$-direction averaged (from the top to bottom panels) stream-wise velocity $\langle{u}\rangle_{yt}$, vertical momentum flux per unit mass $\langle{u'w'}\rangle_{yt}$, stream-wise velocity variance $\langle{u'^{2}}\rangle_{yt}$, vertical velocity variance $\langle{w'^{2}}\rangle_{yt}$ and turbulence kinetic energy per unit mass $\langle{k}\rangle_{yt}$ for \mbox{WRF-R}, compared with the wind-tunnel measurements over the `Furry Hill' from \citet{finnigan95} (see text for details). A graphical representation of the slope in the terrain present around each profile is displayed at the top of the figure}
\label{fig:profiles_rgh}
\end{figure*}


\par
Simulated vertical profiles of the time and $y$-direction averaged stream-wise velocity and turbulence statistics to the counterpart measured profiles are presented in Fig.~\ref{fig:profiles_can} and Fig.~\ref{fig:profiles_rgh} for a selected subset of positions across the ridge. A quantitative evaluation in terms of root-mean-square error (RMSE), denoted by $\gamma$ herein, for all measurement positions $\left({x,h}\right)$, illustrated in Fig.~\ref{fig:guide}, is presented in Table~\ref{tab:rmse1}. Vertical profiles of the time and $y$-direction averaged stream-wise velocity $\langle{u}\rangle_{yt}$ for \mbox{WRF-C} and \mbox{WRF-R} are shown respectively in Fig.~\ref{fig:profiles_can}(a) and Fig.~\ref{fig:profiles_rgh}(a). For \mbox{WRF-C}, these profiles are relatively close to those of the measurements, with similar $\gamma({\langle{u}\rangle_{yt}})$ of $0.9$ to $1.3$\,m\,s$^{-1}$ over the range of horizontal grid spacings. For \mbox{WRF-R} the profiles of $\langle{u}\rangle_{yt}$ are similar to the measurements upstream of the ridge but, as a result of the weaker separation on the lee side of the ridge, there is an excess of $\langle{u}\rangle_{yt}$ from approximately $z/h_c=2$ to $6$. This indicates that using a passive roughness is appropriate to represent the mean boundary-layer flow over flat ground for the case considered, but performance degrades significantly when the boundary-layer flow crosses over the ridge. When evaluated across all measurement positions $\left({x,h}\right)$, $\gamma({\langle{u}\rangle_{yt}})$ is $49$ to $178$\% larger for \mbox{WRF-R} than for \mbox{WRF-C} (see~Table~\ref{tab:rmse1}). The canopy model therefore performs better overall at reproducing the mean flow over a ridge covered in a forest canopy. Although the canopy model in \mbox{WRF-C} was implemented with the \textit{sharp transition} of \citet{ouwersloot17}, erroneous fluctuations in stream-wise velocity above the canopy were not seen.


\begin{table*}
\caption{Domain-wide root-mean-squared-error $\gamma$ for the time and $y$-direction averaged stream-wise velocity $\langle{u}\rangle_{yt}$ (in m\,s$^{-1}$), vertical momentum flux per unit mass $\langle{u'w'}\rangle_{yt}$ (in m$^{2}$\,s$^{-2}$), stream-wise velocity variance $\langle{u'^{2}}\rangle_{yt}$ (in m$^{2}$\,s$^{-2}$), vertical velocity variance $\langle{w'^{2}}\rangle_{yt}$ (in m$^{2}$\,s$^{-2}$) and turbulence kinetic energy per unit mass $\langle{k}\rangle_{yt}$ (in m$^{2}$\,s$^{-2}$), for all cases considered for \mbox{WRF-C} and \mbox{WRF-R}}
\label{tab:rmse1}
\begin{center}
\begin{tabular*}{\textwidth}{c@{\extracolsep{\fill}}ccccc}
\hline
Simulation
      & $\gamma({\langle{u}\rangle_{yt}})$
      & $\gamma({\langle{u'w'}\rangle_{yt}})$
      & $\gamma({\langle{u'^{2}}\rangle_{yt}})$
      & $\gamma({\langle{w'^{2}}\rangle_{yt}})$
      & $\gamma({\langle{k}\rangle_{yt}})$    \\
\hline
WRF-C2 & $0.90$ & $0.70$ & $1.03$ & $0.58$ & $1.20$ \\
WRF-C4 & $0.96$ & $0.67$ & $1.19$ & $0.64$ & $1.24$ \\
WRF-C6 & $1.03$ & $0.56$ & $1.26$ & $0.75$ & $1.13$ \\
WRF-R2 & $2.50$ & $0.68$ & $1.88$ & $1.05$ & $2.08$ \\
WRF-R4 & $1.92$ & $0.62$ & $1.67$ & $0.98$ & $1.69$ \\
WRF-R6 & $1.53$ & $0.57$ & $1.62$ & $1.08$ & $1.45$ \\
\hline
\end{tabular*}
\end{center}
\end{table*}


\par
The simulated $\langle{u}\rangle_{yt}$ on the lee side of the ridge is negative within the canopy for \mbox{WRF-C} and \mbox{WRF-R}, which is not the case for the measurements. However, the measurements of velocity must be interpreted with caution in the region of separated flow, since the crossed hot-wire probes that were used are not appropriate for measuring a reversed flow. Using flow visualisation techniques, \citet{finnigan95} were able to identify a separation region $5.2\,L$ in length. This will be explored in Sect.~\ref{subsec:mean_flow} using cross-sections that show the full extent of the simulated separation region more clearly. \mbox{WRF-C} performs better than \mbox{WRF-R} not only for the mean flow but also for the majority of the turbulence statistics (see rows (b) to (e) in Fig.~\ref{fig:profiles_can} and Fig.~\ref{fig:profiles_rgh} and the second to fifth columns of Table~\ref{tab:rmse1}). The RMSE $\gamma$ for \mbox{WRF-R} are $29$ to $83$\%, $44$ to $81$\% and $28$ to $73$\% larger than that for \mbox{WRF-C} for the stream-wise velocity variance $\langle{u'^{2}}\rangle_{yt}$, vertical velocity variance $\langle{w'^{2}}\rangle_{yt}$ and TKE per unit mass $\langle{k}\rangle_{yt}$, respectively.

\par
In the region upstream of the ridge, the stream-wise velocity variance $\langle{u'^{2}}\rangle_{yt}$ takes a similar form to the measured profiles for both \mbox{WRF-C} and \mbox{WRF-R}, with \mbox{WRF-C} closer to the measured values than \mbox{WRF-R} and \mbox{WRF-R2} performing particularly poorly above the canopy. The vertical velocity variance $\langle{w'^{2}}\rangle_{yt}$ and vertical momentum flux per unit mass $\langle{u'w'}\rangle_{yt}$ below $z/h_c=5$ in the upstream region are minimal for all cases for \mbox{WRF-R} in comparison to the corresponding measurements. This is a clear indication that the roughness-length approach to modelling the effects of a canopy does not generate the level of turbulence seen in the wind-tunnel measurements, at least over flat ground. Conversely, \mbox{WRF-C} is able to reproduce most turbulence statistics accurately, with $\langle{u'^{2}}\rangle_{yt}$ and $\langle{k}\rangle_{yt}$ agreeing well with the measured profiles. The magnitude of the peak in $\langle{w'^{2}}\rangle_{yt}$ is reproduced well by \mbox{WRF-C2}; it is between $4$ and $16$\% of the measured peak value in the upstream region. However, the simulations with larger horizontal grid spacings under-predict $\langle{w'^{2}}\rangle_{yt}$, with peak values in this region $22$ to $37$\% smaller for \mbox{WRF-C4} and $40$ to $69$\% smaller for \mbox{WRF-C6}. The \mbox{WRF-C} simulations tend to over-predict the magnitude of $\langle{u'w'}\rangle_{yt}$ above the canopy, with \mbox{WRF-C6} performing slightly better than \mbox{WRF-C4} and \mbox{WRF-C2} in the upstream region.

\par
As the flow reaches the top of the ridge, at $x/L=0$, a peak forms in the measured vertical momentum flux per unit mass $\langle{u'w'}\rangle_{yt}$ within close proximity of the top of the canopy. This is reproduced well by \mbox{WRF-C2}, but \mbox{WRF-C4} and \mbox{WRF-C6} under-predict this peak value respectively by $36$ and $56$\%. For \mbox{WRF-R} this peak is not present for any of the horizontal grid spacings used. A similar peak in TKE per unit mass $\langle{k}\rangle_{yt}$ is also seen in the measurements at this location. In carrying out simulations similar to those discussed here, \citet{dupont08b} and \citet{ross05} reported large peaks in $\langle{k}\rangle_{yt}$ near the displacement height of the canopy at ridge-top when using a change in roughness length at the surface to represent the canopy. While this was not seen in the results shown here, such an excess was present in the results of preliminary simulations carried out by the authors when considering non-neutral conditions. The peaks in $\langle{k}\rangle_{yt}$ for \mbox{WRF-R} at this location are approximately $50$\% smaller than the measured values, with \mbox{WRF-R6} and \mbox{WRF-R4} $11$ to $12$\% closer to the measured peak value than \mbox{WRF-R2}. The peak values of $\langle{k}\rangle_{yt}$ for \mbox{WRF-C} are within $12$ and $23$\% of the measured peak values, with smaller horizontal grid spacings providing the closest agreement. The differences between the ridge-top profiles for the different grid-spacings are minimal at most locations, but while \mbox{WRF-C2} provides the best result for the canopy model simulations, \mbox{WRF-R6} provides the best result for the simulations using only a change in roughness length at the surface.

\par
As the flow proceeds downstream, all turbulence statistics begin to respond more strongly to the presence of the forested ridge. For the simulations and measurements, the momentum flux $\langle{u'w'}\rangle_{yt}$ displays a single peak, the location of which increases in height above ground level between $x/L=0$ and $2.5$ (see Fig.~\ref{fig:profiles_can}(b) and Fig.~\ref{fig:profiles_rgh}(b)). It should be noted that these figures show height above ground level. Thus, while a peak in a profile appears to be displaced upwards, it actually remains at a fairly constant altitude, as will be shown in the cross-sections of Sect.~\ref{subsec:mean_flow}.

\par
For \mbox{WRF-C}, the peak in $\langle{u'^{2}}\rangle_{yt}$ increases in height downstream over a similar range of $x/L$ to that seen in the measurements. However, for $\langle{w'^{2}}\rangle_{yt}$ the peak rises initially but then remains between $h/h_c=1.5$ to $2.5$ over $x/L=0.7$ to $2.5$. $\langle{u'w'}\rangle_{yt}$ and $\langle{k}\rangle_{yt}$ correspondingly increase more rapidly at low levels than higher above the ground. This leads to the peak value occurring lower than that measured in the wind tunnel and to closer agreement with the measurements above the peak than below. For \mbox{WRF-R} the peak value of $\langle{w'^{2}}\rangle_{yt}$ occurs slightly closer to the ground, between $h/h_c=1$ to $2$. The magnitude of this peak is not reproduced well by any of the \mbox{WRF-R} simulations at $x/L=0.7$. For $x/L=1$ to $1.4$, \mbox{WRF-R6} and \mbox{WRF-R4} remain close to the measured profiles of $\langle{w'^{2}}\rangle_{yt}$ until $h/h_c=1.5$ but under-predict above this height until $h/h_c=6$. This is also true for the peak in $\langle{w'^{2}}\rangle_{yt}$ for \mbox{WRF-R2}, except by $x/L=1.4$ the peak value is over-predicted by $32$\% while still $1.5\,h_{c}$ lower than the height of the peak in the measurements. The peaks in the downstream profiles of $\langle{u'^{2}}\rangle_{yt}$ for \mbox{WRF-R} rise over a range of $x/L$ more in line with the measurements and \mbox{WRF-C} than those for $\langle{w'^{2}}\rangle_{yt}$. However, by $x/L=1.4$ these peaks are $0.5$ to $1\,h_{c}$ lower than those in the measurements, with this discrepancy most acute in the case of \mbox{WRF-R2}. This leads to the peak in $\langle{u'w'}\rangle_{yt}$ for \mbox{WRF-R} occurring up to $1.5\,h_{c}$ lower than the peak in the measurements, compared to $1\,h_{c}$ lower for \mbox{WRF-C}. As \mbox{WRF-R} reproduces $\langle{u'w'}\rangle_{yt}$ more accurately downstream of the ridge and less accurately before the ridge and {\it vice versa} for \mbox{WRF-C}, the corresponding $\gamma$ for this quantity are within $2$ to $8$\% when comparing the two methods of parametrising the canopy.

\par
The results produced by \mbox{WRF-C} are very similar to those that were produced in the simulations carried out by \citet{dupont08b}. The peak in $\langle{u'w'}\rangle_{yt}$ immediately downstream of the ridge also occurs at a position in the vertical that is different from that of the measurements and is over-predicted by a similar amount, but by $x/L=3.5$ the profile is more similar to the measurements than \mbox{WRF-C}. The simulations from both works follow the profiles of $\langle{u'^{2}}\rangle_{yt}$ and $\langle{k}\rangle_{yt}$ closely. Similar but small over-predictions in these quantities below $z/h_c=3$ are present in both cases, but larger for \mbox{WRF-C}. While \mbox{WRF-C2} performs considerably better in reproducing $\langle{w'^{2}}\rangle_{yt}$ upstream of the ridge, the results by \citet{dupont08b} are closer to the measurements downstream of the ridge. The vertical position of the peak in $\langle{w'^{2}}\rangle_{yt}$ is also lower than in the measurements at $x/L=2$ but follows the measurements better than any of the \mbox{WRF-C} simulations. It is worth noting that \citet{dupont08b} do not provide the value for $u_\star$ used for normalisation and this could lead to differences in the magnitudes of the various statistics between those results and the results of \mbox{WRF-C}. The similarity between the height of the wake seen in the \mbox{WRF-C} simulations and those of \citet{dupont08b} would suggest that the difference to the measurements is a result of the experiment being scaled up. It is therefore possible that further similarity conditions are required when scaling up experiments studying flows over ridges covered with a canopy. The `Furry Hill' experiment was studied by \citet{ross05} using numerical simulations of the same scale as that of the wind tunnel with a canopy model and using only a change in roughness length at the surface to parametrise the artificial canopy. However, it is difficult to compare \mbox{WRF-C} and \mbox{WRF-R} to those results due to the reduced vertical extent of the plotted data and the small size of the plots themselves.

\par
There are some considerable differences between the profiles of turbulence statistics for the equivalent simulations with different horizontal grid spacings. Upstream of the ridge, \mbox{WRF-C2} provides the closest agreement with the measurements but, as the flow proceeds past the ridge, \mbox{WRF-C6} tends to produce better results. The finest grid appears to be reproducing the fine scales of turbulence above the flat section of canopy upstream of the ridge well, but is not performing so well in the wake, where turbulence is generated by the combination of the canopy and the ridge. The profiles of turbulence statistics for \mbox{WRF-R6} and \mbox{WRF-R4} are very similar across most positions, although with considerable differences to the measured values. However, there are large differences between the profiles for \mbox{WRF-R2} and those with a more coarse grid spacing. The TKE $\langle{k}\rangle_{yt}$ and vertical momentum flux $\langle{u'w'}\rangle_{yt}$ are over-predicted to varying degrees in all of the simulations below $h/h_c=5$ to $6$ in the downstream region. While this is still true at $x/L=3.5$, the differences between equivalent simulations of different horizontal grid spacings are greatly reduced, with profiles of $\langle{k}\rangle_{yt}$ much closer to the measured profiles. The horizontal grid resolution resolution of the simulations has a strong influence on the properties of the flow in close proximity to the forested ridge, but does not make a large difference to the properties of the flow further downstream.

\par
While the magnitude of the turbulence statistics generated from \mbox{WRF-R} and \mbox{WRF-C} both differ from the measurements in some positions, the forms of the vertical profiles are more closely reproduced by \mbox{WRF-C}. The response to the forested ridge for \mbox{WRF-R} generally lies between what would be expected for a ridge with negligible roughness and that for \mbox{WRF-C}. The roughness-length approach to modelling the effects of a canopy does modify the dynamics correctly but not sufficiently for the extensive, tall canopy considered in the present work. In sum, the comparison of model results with the `Furry Hill' data shows that the implementation of the canopy model in WRF is a significant improvement over the roughness-length approach for both the mean flow and the turbulence statistics when considering a ridge covered by a tall canopy for the range of horizontal grid spacings considered herein. While \mbox{WRF-C} provides better results than \mbox{WRF-R} immediately downstream of the ridge, there are still clearly deficiencies in the model's ability to reproduce the magnitude and height of the wake.


\subsection{Flow Features}
\label{subsec:mean_flow}


\begin{figure*}
\centering{\scalebox{1.0}{\includeCroppedPdf[width=1.0\linewidth]{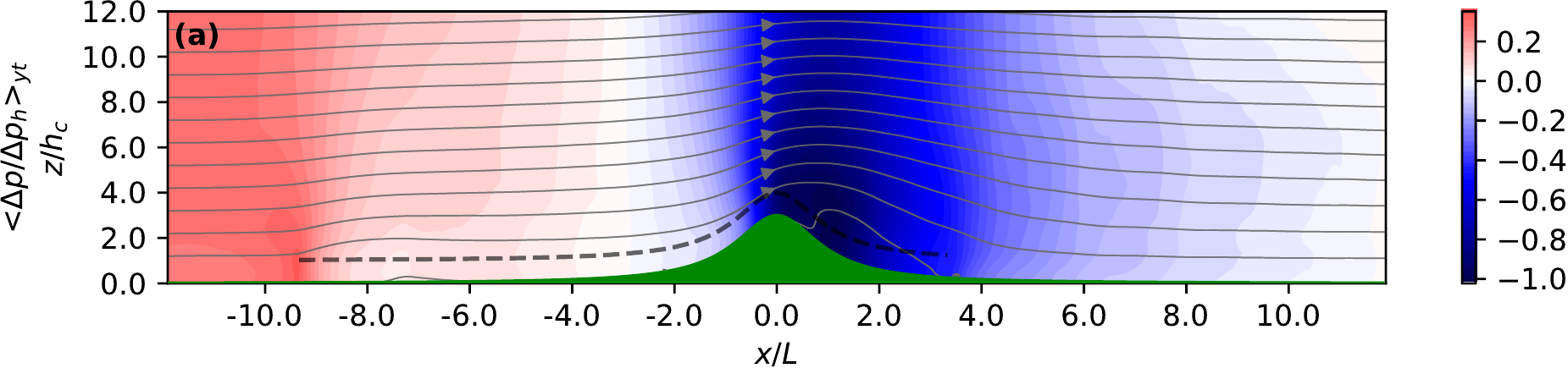}}}
\centering{\scalebox{1.0}{\includeCroppedPdf[width=1.0\linewidth]{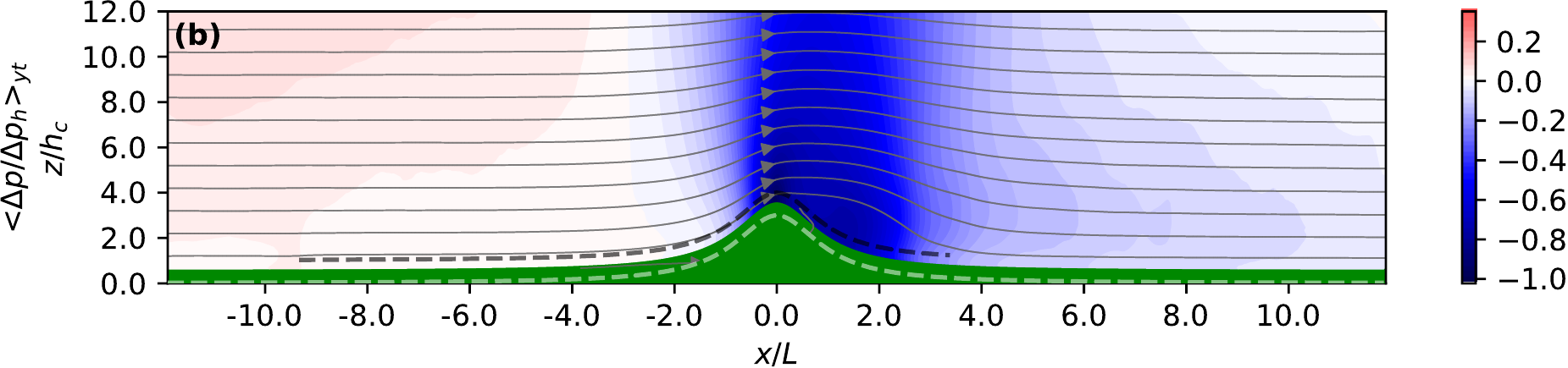}}}
\caption{Simulated mean normalised pressure perturbation $\langle{\Delta{p}}\rangle_{yt}$ for {\bf (a)} \mbox{WRF-C4} and {\bf (b)} \mbox{WRF-R4}. The pressure perturbation is calculated relative to the value at the corresponding height at $x/L=-3$. The top of the simulated artificial canopy is indicated by a black dashed line. The light grey lines show mean flow streamlines originating at regular height intervals above the flat ground at $x/L=-12$. The dashed white line in {\bf (b)} represents the effective ground level of the vertically displaced \mbox{WRF-R4} simulation (see Sect.~\ref{sec:model_setup} for details)}
\label{fig:cross_p}
\end{figure*}


Using the scaling arguments presented at the beginning of Sect.~\ref{subsec:evaluation}, since $h_{c}/L_{c}\sim H/L$ in the present work, there ought to be an interplay between the canopy and the ridge on the generation of drag on the ridge surface. The mean pressure perturbation field $\langle{\Delta{p}}\rangle_{yt}\left({x,z}\right)$ across the finer-resolved domain is shown in Fig.~\ref{fig:cross_p} for \mbox{WRF-C4} and \mbox{WRF-R4}. Significant differences can be noticed between the two simulations, most notably the presence of a local maximum and minimum in pressure respectively before and after the leading edge of the canopy at $x/L=-9.35$ for \mbox{WRF-C} (see Fig.~\ref{fig:cross_p}a). The local minimum at $x/L=-7.5$ induces an adverse pressure gradient, thereby decelerating the flow \citep[see][for a detailed description of the adjustment of a turbulent boundary layer to a canopy of roughness elements]{belcher03}. In contrast, pressure is essentially horizontally uniform upstream of the ridge for \mbox{WRF-R} with only a slight decrease caused by the change in roughness length at $x/L=-9.35$. The flow is assumed to adjust to the canopy at the location where $\langle{w}\rangle_{yt}/u_{\mbox{\scriptsize ref}}=0.01$, with $u_{\mbox{\scriptsize ref}}$ taken as $\langle{u}\rangle_{yt}$ at height $h/h_{c}=2$ over the flat section of terrain with no canopy present at $x/L=-10.5$. For all cases of \mbox{WRF-C} this adjustment length $L_{a}\approx3.8\,L_{c}$, which is smaller than the range $4.5$--$6\,L_{c}$ predicted by \citet{belcher12}. However, it is in line with the values of $3\,L_{c}$ and $4\,L_{c}$ found in the analytical and numerical studies of \citet{belcher08} and \citet{dupont09}, respectively. It should be noted that the analytical methods use the location where the vertical velocity has dropped to the friction velocity $u_{\star}$; however for \mbox{WRF-C} the vertical velocity is always less than the friction velocity at canopy top.

\par
Pressure decreases and hence the wind speed increases as the flow approaches the top of the ridge. In the case of a ridge with no canopy and negligible surface roughness, the pressure minimum is located directly above the ridge. When a canopy is present on the ridge this pressure minimum is displaced to a position downstream from the ridge-top (see Fig.~\ref{fig:cross_p}). For \mbox{WRF-C} the area of lowest pressure extends over the majority of the slope on the lee side of the ridge for all horizontal grid spacings. A pressure minimum occurs immediately downstream of the ridge-top just above the canopy at $x/L=0.95$, $h/h_{c}=1.7$ for \mbox{WRF-C4}. Changing the horizontal grid spacing does not modify the height of this minimum by more than $0.1\,h_{c}$. However larger grid spacings result in a more significant displacement of this minimum in the stream-wise direction, $x/L=0.67$ for \mbox{WRF-C2} and $x/L=1.14$ for \mbox{WRF-C6} (not shown). While there is an area of reduced pressure over a similar extent for \mbox{WRF-R}, the location of minimum pressure is at the top of the ridge at the lowest modelled level at $x/L=0$, $h/h_{c}=0.6$ for all horizontal grid spacings considered.


\begin{figure*}
\centering{\scalebox{1.0}{\includeCroppedPdf[width=1.0\linewidth]{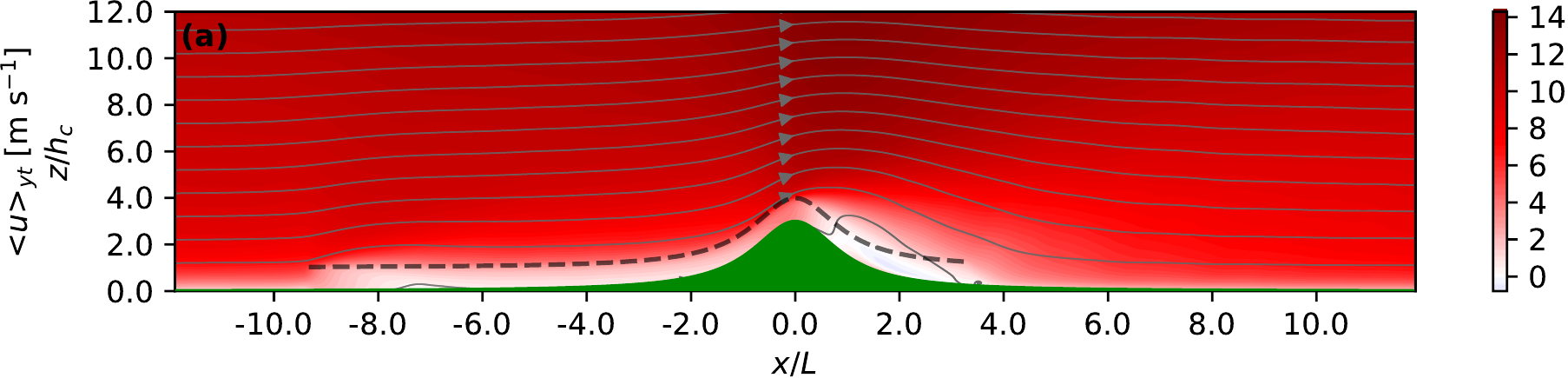}}}
\centering{\scalebox{1.0}{\includeCroppedPdf[width=1.0\linewidth]{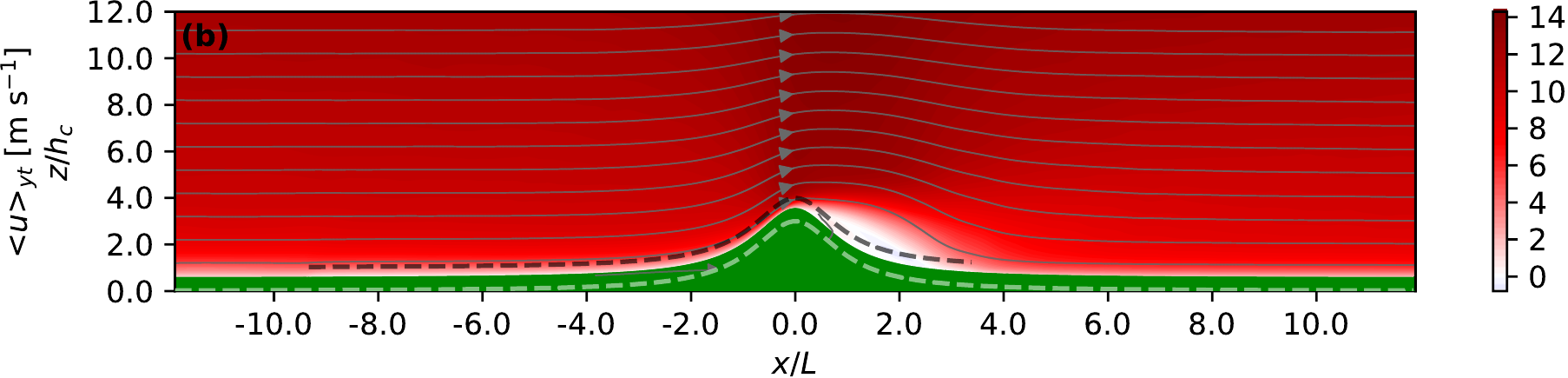}}}
\caption{Simulated mean stream-wise velocity $\langle{u}\rangle_{yt}$ for {\bf (a)} \mbox{WRF-C4} and {\bf (b)} \mbox{WRF-R4}. The top of the simulated artificial canopy is indicated by a black dashed line. The light grey lines show mean flow streamlines originating at regular height intervals above the flat ground at $x/L=-12$. The dashed white line in {\bf (b)} represents the effective ground level of the vertically displaced \mbox{WRF-R4} simulation (see Sect.~\ref{sec:model_setup} for details)}
\label{fig:cross_u}
\end{figure*}


\par
On the lee side of the ridge the adverse pressure gradient leads to flow separation which, in turn, causes the adverse pressure gradient to extend further downstream as if the ridge has been extended in the downstream direction. This effect is more significant for \mbox{WRF-C} than for \mbox{WRF-R}, although the pressure field for \mbox{WRF-R} is much closer to that for \mbox{WRF-C} than to that which would be expected for a ridge with negligible roughness. The pressure close to the ground readjusts downstream of the ridge over a shorter distance for \mbox{WRF-R} than for \mbox{WRF-C} (cf.~Fig.~\ref{fig:pressure1}). Coupling between the out-of-phase flows within and above the canopy results in a reduced pressure gradient (reduced over-speeding) over the ridge for \mbox{WRF-C} compared with that for \mbox{WRF-R}. The separation region extends over $4\,L$ in length over the ground surface for all cases for \mbox{WRF-C}, in line with the experimental data of \citet{finnigan95} and numerical data of \citet{ross05} and \citet{dupont08b}, compared with $3\,L$ for \mbox{WRF-R4} (see Fig.~\ref{fig:cross_u}). However, the horizontal extent of the separation region at the surface for \mbox{WRF-R} would be of comparable extent if the flow could be visualised below the displacement height $d$. The horizontal grid spacing does not modify the length of the separation region significantly for \mbox{WRF-C}. For \mbox{WRF-R} the horizontal extent of the separation region increases with horizontal grid spacing, from $2.2\,L$ in length for \mbox{WRF-R2} to $3.5\,L$ for \mbox{WRF-R6}.


\begin{figure*}
\centering{\scalebox{1.0}{\includeCroppedPdf[width=1.0\linewidth]{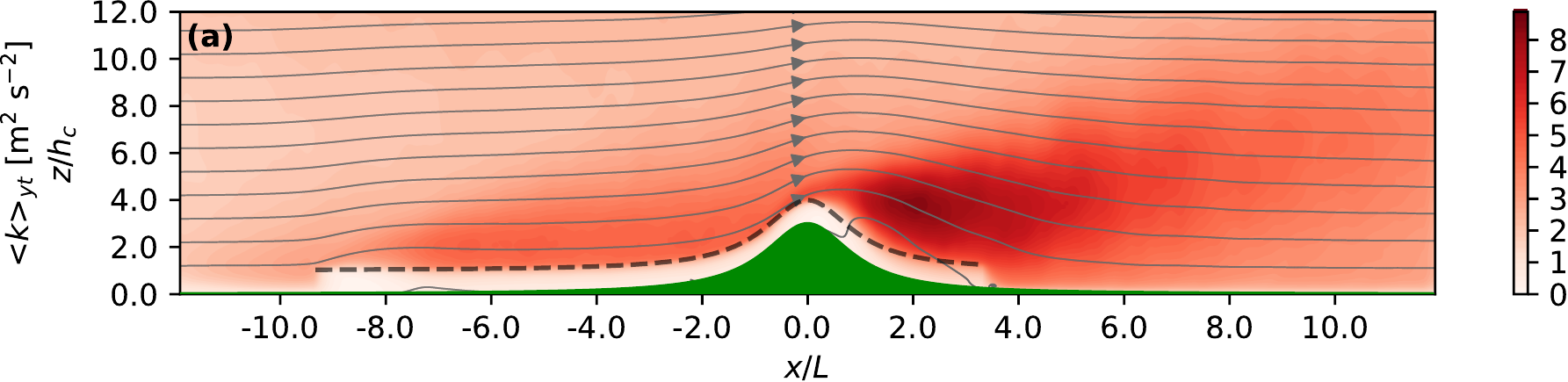}}}
\centering{\scalebox{1.0}{\includeCroppedPdf[width=1.0\linewidth]{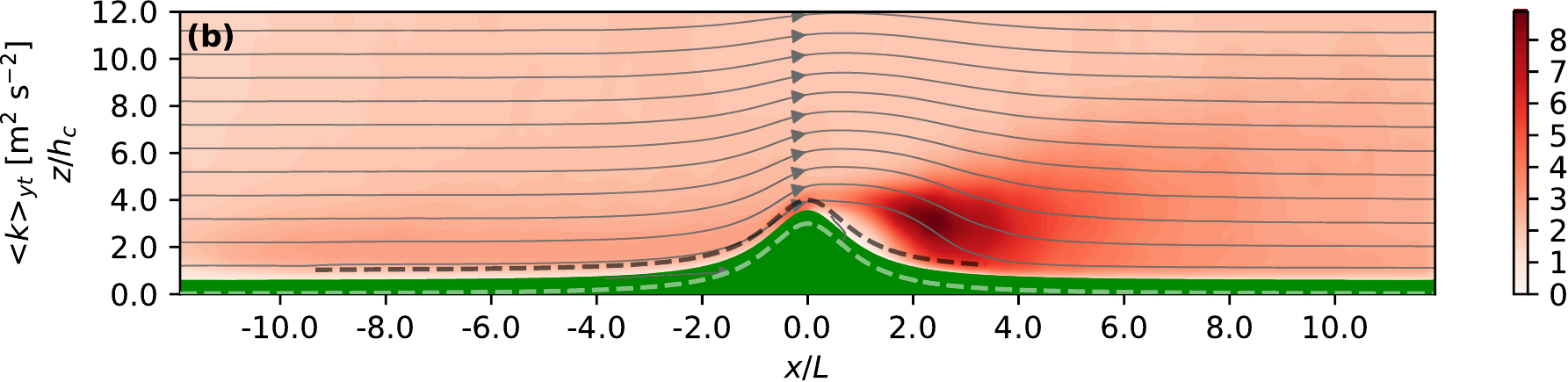}}}
\caption{Simulated turbulence kinetic energy per unit mass $\langle{k}\rangle_{yt}$ for {\bf (a)} \mbox{WRF-C4} and {\bf (b)} \mbox{WRF-R4}. The top of the simulated artificial canopy is indicated by a black dashed line. The light grey lines show mean flow streamlines originating at regular height intervals above the flat ground at $x/L=-12$. The dashed white line in {\bf (b)} represents the effective ground level of the vertically displaced \mbox{WRF-R4} simulation (see Sect.~\ref{sec:model_setup} for details)}
\label{fig:cross_k}
\end{figure*}


\par
A wake is created on the lee side of the ridge, centred vertically on the region of maximum wind shear above the separation region as shown in Fig.~\ref{fig:cross_k} for $\langle{k}\rangle_{yt}$ (although a similar structure is visible in all other turbulence statistics). The vertical differential in wind speed at the top of the ridge is smaller for \mbox{WRF-R} than that at canopy height at the top of the ridge for \mbox{WRF-C}. Therefore, the wake angle and the intensity of turbulence within the wake are larger for \mbox{WRF-C} than for \mbox{WRF-R}. There is evidence of Kelvin-Helmholtz billows forming in the wake for both simulations. Turbulence is suppressed in the canopy for \mbox{WRF-C} while fluctuations are clearly visible near the surface for \mbox{WRF-R}. The vertical spread or depth $D$ of the turbulent wake region follows a power law of the form $D=A\,\left({x-x_{0}}\right)^{\alpha}$, as presented by \citet{kaimal94} in reference to \citet{taylor88}. In this formulation $x_{0}$ is a virtual origin situated before the ridge and $A$ is a constant. \citet{kaimal94} pointed out that theory, wind-tunnel and field experiments have not decided whether $\alpha$ should be equal to $0.5$ or $1$. The power $\alpha$ that best fit the wakes in the simulations presented here is in the range $0.6-0.7$ with $x_{0}$ taken as $x/L=-3$.


\section{Conclusions and Discussion}
\label{sec:conc}

Results from numerical model simulations of neutral boundary-layer flow across a forested ridge using a canopy model (\mbox{WRF-C}) or a bare surface with an increased roughness $z_{0}$ at the location of the canopy (\mbox{WRF-R}) using a range of resolutions were analysed and compared. The main conclusions, along with some discussion, are given below.

\begin{itemize}
\item[$\bullet$]{The speed of the flow in the stream-wise direction is closer to the counterpart wind-tunnel measurements for \mbox{WRF-C} than for \mbox{WRF-R}. As is expected, the reduced canopy drag for \mbox{WRF-R} leads to an over-estimation of the wind speed above the canopy, which becomes larger as the flow proceeds downstream.}
\item[$\bullet$]{\mbox{WRF-C} captures the measured turbulence statistics significantly better than \mbox{WRF-R}. The boundary layer has very little turbulence upstream of the ridge for \mbox{WRF-R} with a particular deficiency in the vertical velocity variance $\langle{w'^{2}}\rangle_{yt}$.}
\item[$\bullet$]{While the forested ridge in \mbox{WRF-R} generates turbulence close to the ground, the vertical extent of these turbulent structures does not reach as far above the ground as those seen in the measurements or in \mbox{WRF-C}.}
\item[$\bullet$]{For \mbox{WRF-R} the horizontal extent of the separation region increases as the horizontal grid spacing is increased. This is not seen in \mbox{WRF-C}, where the separation region was of comparable extent for each horizontal grid spacing considered.}
\item[$\bullet$]{The discrepancies between the experimental measurements and simulated values of stream-wise velocity and the various turbulence statistics are reduced by reducing the horizontal grid spacing for \mbox{WRF-C}.}
\item[$\bullet$]{While it might be expected for a finer horizontal resolution to improve the results of \mbox{WRF-R}, the discrepancy with the measurements is actually increased, as is discussed below.}
\end{itemize}

\par
The RMSE $\gamma$ between the measured and modelled profiles at the positions shown in Fig.~\ref{fig:guide} varies between the different turbulence statistics and mean stream-wise velocity for different horizontal grid spacings for \mbox{WRF-C} and \mbox{WRF-R}. In general, \mbox{WRF-C} provides closer results to the measurements when a smaller horizontal grid spacing is used, while the \mbox{WRF-R} simulations provide closer results at larger grid spacings for the case considered here. The RMSE for vertical momentum flux $\gamma({\langle{u'w'}\rangle_{yt}})$ and TKE $\gamma({\langle{k}\rangle_{yt}})$ do not follow this trend for \mbox{WRF-C}, with the largest grid spacing providing the best result in the reproduction of both statistics. However, this discrepancy is predominantly due to the turbulence occurring directly after the peak of the ridge at the profiles between $x/L=0.7$ and $1.4$. In this region there is a large over-prediction in $\langle{w'^{2}}\rangle_{yt}$ at low levels and the closest simulated profiles to the measurements shift from \mbox{WRF-C2} to \mbox{WRF-C6} as the flow moves downstream. The stronger response to the ridge at low levels when smaller horizontal grid spacings are used leads to an even greater over-prediction in $\langle{u'w'}\rangle_{yt}$ and, as the difference to the measurements is large here, $\gamma({\langle{u'w'}\rangle_{yt}})$ is heavily influenced by these profiles. The smaller grid spacings also lead to an over-prediction of $\langle{k}\rangle_{yt}$ in this region, but the difference this makes to $\gamma({\langle{k}\rangle_{yt}})$ is reduced by the smaller difference in magnitude to the measured profiles. The fact that \mbox{WRF-C2} is closer to the measurements before and after the ridge leads to $\gamma({\langle{k}\rangle_{yt}})$ for \mbox{WRF-C2} being only $6$\% larger than for \mbox{WRF-C6}, while for $\gamma({\langle{u'w'}\rangle_{yt}})$ \mbox{WRF-C2} is $25$\% larger. For \mbox{WRF-R}, $\gamma({\langle{u'w'}\rangle_{yt}})$ and $\gamma({\langle{k}\rangle_{yt}})$ reduce with increasing horizontal grid spacing in line with the other statistics. However, $\langle{w'^{2}}\rangle_{yt}$ is reproduced poorly for all grid spacings, with $\gamma({\langle{w'^{2}}\rangle_{yt}})$ remaining at approximately $1$\,m$^{2}$\,s$^{-2}$.

\par
The larger discrepancies between the measurements and the counterpart numerical results, when smaller horizontal grid spacings are used, is also likely due to a compounding of errors. The dynamics of the flow in and around the canopy is not properly modelled when the canopy is represented only by a change in roughness length at the surface, as for \mbox{WRF-R}. When the horizontal grid spacing is reduced there are a larger number of grid cells over which errors can accumulate. For \mbox{WRF-C} the flow is more accurately reproduced over the flat terrain for which the canopy model used was devised. However, immediately downstream of the ridge, the smallest grid spacing is not providing a significant improvement to the reproduction of turbulence in the wake region. This may indicate that the canopy model used for \mbox{WRF-C} requires improvement to properly reproduce canopy dynamics in complex terrain. However, this canopy model is representing an evenly spaced array of cylindrical stalks as a homogeneous, porous block and so inconsistencies are always likely to be present.

\par
For each of the simulations performed here, with a vertical resolution at the surface of $0.1\,h_c$, a corresponding simulation was carried out with a vertical resolution of $0.2\,h_c$. These have not been shown here as the differences between these two sets of simulations were negligible at all positions, other than a shift of approximately $z/h_c=0.1$ in the height of peak values at canopy top due to the larger vertical extent of the grid cells. When a forest canopy is represented using an increase in roughness length at the surface and vertical resolution smaller than that of canopy elements, it is necessary to elevate the ground to the displacement height of the canopy. As this is not practical in numerical simulations, it is difficult to accurately reproduce the modification to the flow by the leading or trailing edge of the canopy using such a method. An additional simulation was performed with a bottom grid cell height of $h/h_c=1.2$, using the roughness length approach but without displacing the ground. Results were found to have the same deficiencies as the other \mbox{WRF-R} simulations, with little to no response to the canopy visible in the turbulence statistics (not shown). Hence, a change in roughness length cannot replicate the effects of a relatively tall canopy in the modulation of the flow speed over a ridge. An explicit treatment of the canopy represents a significant improvement over the roughness-length approach. However, it does require a fine spatial resolution in the vertical and the horizontal to include model layers within the canopy and to model turbulence in a large-eddy simulation mode. The high spatial resolution of the simulation leads to steeper slopes in the orography, which set constraints on the time-step \citep{Connolly20}. Further research will be required to determine the effectiveness of this canopy model if coarser spatial resolutions are required.

\par
While the results presented herein are for a fairly idealised terrain geometry and set of initial and boundary conditions, the results have important practical implications for the assessment and management of wind and pollution in the atmosphere within complex terrain. In the interest of comparing to the measurements of \citet{finnigan95}, the current study was limited to geometry of the hill and canopy in their experiment with $L_c/L=0.36$ and $h_c/H=0.33$. For this case it appears that a horizontal resolution of $0.024\,L$ ($0.066\,L_c$) and a vertical resolution of $0.1$ to $0.2\,h_c$ was appropriate to reproduce the flow over a forested ridge using a canopy model with vertical extent. Further work is required to explore a wider range of hill geometries and canopy properties and extents at different resolutions, as well as real-case studies that are likely to demand coarser grid resolutions. However, this will require considerably more experimental data on canopy flows in complex terrain to be collected.


\begin{acknowledgements}
The authors thank Dr. Sylvain Dupont for the provision of data from the paper by \citet{dupont08b}. Numerical model simulations were performed using the University of Hertfordshire high-performance computing facility.
\end{acknowledgements}


\bibliographystyle{spbasic}


\end{document}
